\documentclass[preprint,12pt,authoryear]{elsarticle}

\usepackage[left]{lineno}  

\usepackage{graphicx}
\usepackage{newtxtext}
\usepackage{newtxmath}
\usepackage{natbib}
\usepackage{hyperref}
\hypersetup{
    colorlinks = true,
    urlcolor   = blue,
    citecolor  = black,
}

\newcommand{\RomanNumeralCaps}[1]
\linenumbers

\usepackage{natbib}
\usepackage{todonotes}
\usepackage{booktabs}

\usepackage{dsfont}

\usepackage{float}

\newcommand*\xbar[1]{%
   \hbox{%
     \vbox{%
       \hrule height 0.5pt 
       \kern0.5ex
       \hbox{%
         \kern-0.1em
         \ensuremath{#1}%
         \kern-0.1em
       }%
     }%
   }%
}

\usepackage{url}
\usepackage{amsmath}

\setcitestyle{citesep={,}}

\usepackage{psfrag}
\usepackage{epsfig}
\usepackage[caption=false]{subfig}
\usepackage{graphicx}
\usepackage{upgreek}

\usepackage{titlesec}

\usepackage[legalpaper, margin=1in]{geometry}

\journal{Chemical Engineering Journal}

\begin{document}

\begin{frontmatter}



\title{Stationary states of forced two-phase turbulence} 


\author[label1,label2]{Suhas S. Jain} 
\ead{suhasjain@gatech.edu, Corresponding Author}

\author[label2]{Ahmed Elnahhas} 

\affiliation[label1]{organization={Flow Physics and Computational Science Lab, George W. Woodruff School of Mechanical Engineering, Georgia Institute of Technology},
            state={GA},
            country={USA}}

\affiliation[label2]{organization={Center for Turbulence Research, Stanford University},
            state={CA},
            country={USA}}

\begin{abstract}
In this work, we perform numerical simulations of forced two-phase isotropic turbulence to study the stationary states of a two-phase mixture.  
We first formulate three different approaches to force a two-phase turbulent flow that maintains a constant (a) mixture kinetic energy, (b) mixture kinetic energy + surface energy, and (c) individual phase kinetic energy, and study their effect on the final stationary state of the system. We show that forcing the two-phase system eliminates the arbitrariness associated with the initialization of the second phase in two-phase turbulence simulations. 
Next, we study the global statistics of the two-phase mixture in the stationary state for varying density ratios ($10^{-3}$ to $10^3$) and viscosity ratios ($10^{-3}$ to $10^3$), void fraction (dilute to dense regimes), and Weber numbers (breakup and non-breakup regimes). 
We utilize a spatial filtering approach, that is applicable for the general case of incompressible/compressible two-phase flows, to compute the energy spectra of each of the phases and study the turbulent kinetic energy distribution across scales. We find that the interface behaves as a soft wall, inhibiting the accommodation of eddies inside the dispersed phase.
Finally, we show that bubbles and drops exhibit different turbulent characteristics for the same global conditions of the carrier phase, with bubbles exhibiting higher turbulent intensities inside them and droplets behaving ballistically, similar to solid particles. 
This study acts as a foundational work on stationary two-phase turbulence, which can be used for further analysis and the development of subgrid-scale models for larger-scale simulations of two-phase turbulent flows.

\end{abstract}




\end{frontmatter}



\section{Introduction}

The deformation, fragmentation, and coalescence of bubbles and droplets in a turbulent environment is an important problem in various environmental and industrial settings \citep{liao2009literature,liao2010literature,jain2017flow,ni2024deformation}. In the atmosphere, the enhanced efficiency of water droplet collision\textendash coalescence events due to air turbulence heavily influences cloud formation and warm rain initiation \citep{grabowski2013growth}. At the ocean\textendash atmosphere interface, underwater turbulence generated due to breaking waves breaks apart large entrained air pockets into smaller bubbles, facilitating mass transfer between the ocean and the atmosphere \citep{deike2022mass}. Finally, turbulent emulsions, i.e., a homogenized mixture of two fluids of similar densities, are ubiquitously formed for use in pharmaceutical and industrial processing \citep{wang2007oil,kilpatrick2012water}. As such, an understanding of how turbulence influences (and is influenced by) the dispersed phase, as parameters such as density and viscosity ratios, between the carrier and dispersed phases, as well as the volume fraction occupied by the dispersed phase are varied, is of paramount importance to understanding these ubiquitous flows. 

While these examples contain irreducible effects due to anisotropy, large-scale shear \citep{rosti2019droplets}, and buoyancy \citep{ravelet2011dynamics}, the most straightforward setup to study the essential physics of deformation, fragmentation, and coalescence of bubbles and droplets in a turbulent environment is homogeneous isotropic turbulence (HIT). While simplistic, this setup allows for systematically studying the high-dimensional parameter space of density and viscosity ratios, volume fraction of dispersed phase, and Weber and Reynolds
numbers. For each of these global parameters, the dependence of the phenomenology of the flow on the scales of turbulent fluctuations in the carrier phase and the scales of the dispersed phase can be further probed both above and below the Hinze scale \citep{hinze1955fundamentals} (scale at which the inertial forces causing fragmentation due to turbulence are balanced by the restorative forces of surface tension). Incorporating the appropriate phenomenology of fragmentation and coalescence as a function of the global and scale-dependent parameters is essential to develop coarse-grained statistical models that can accurately capture the quantitative properties of the mixture. For example, while similar density mixtures lead to large-scale filamentary deformations requiring capillary-like breakup models \citep{eastwood2004breakup,kim2020subgrid}, large density ratios such as in the case of air-water bubble break-up might be better modeled using a cascade-like model that may or may not account for the intermittent nature of turbulence \citep{chan2021turbulenta,chan2021turbulentb,qi2022fragmentation}.

Recent studies that utilize two-phase HIT as a testbed can be broken down into two categories: (i) The first category considers the detailed dynamics of individual bubbles and droplets, or emulsions,
during the first few breakup events through simulations containing a single bubble/droplet, characterizing and quantifying things such as the generation of sub-Hinze scale bubbles due to rapid successive breakup events and capillary fragmentation \citep{riviere2021sub,riviere2022capillary}, the effect of eddies far away from the two-phase interface on the deformation process \citep{vela2021deformation}, bubble deformation modes due to turbulent fluctuations \citep{perrard2021bubble}, the non-vanishing probability of breakup at sub-Hinze scales due to increasingly intermittent turbulent fluctuations \citep{vela2022memoryless}, and the role of the viscosity ratio between the two phases on the rate of bubble breakup and the critical Weber number at which they break \citep{farsoiya2023role}. 
(ii) The second category considers characterizing the statistical properties of numerous bubbles and droplets present simultaneously in a turbulent environment. Within this category, \citet{dodd2016interaction} studied the turbulent kinetic energy (TKE) budgets of droplets at various density ratios within decaying HIT. \citet{crialesi2022modulation, crialesi2023interaction} studied the influence of viscosity ratios and volume fractions of the dispersed phase at unity density ratios in a forced HIT setting on the scale-by-scale energy budgets while linking the droplet-size distribution (DSD) to the local dissipation rate of turbulence, once again highlighting the importance of intermittency in the breakup dynamics. Finally, \citet{begemann2022effect} and \citet{krzeczek2023effect} utilize a linear forcing approach to study the segregation properties of emulsions, highlighting that even minor variations in the density ratio of the two phases from slightly below to slightly above unity shifts the DSD to have fewer super-Hinze bubbles and more sub-Hinze scale drops.  

In both of the aforementioned categories of studies, decaying turbulence cases are utilized considerably. However, this dependence on decaying turbulence can introduce two potential hidden influences on the statistical property of interest. The first concerns the initialization process of the dispersed phase. For example, the initial velocity fields inside the dispersed phase are most often, and somewhat arbitrarily, set as zero. It is unclear whether this affects the fragmentation and coalescence statistics under study, e.g., critical Weber numbers, since no systematic study of the effect of the internal initial condition on the breakup statistics has been performed. In this paper, we will show that the effect of the initial condition lasts for a considerable amount of time on quantities of interest such as the total surface area, and hence, measurements taken during times before the first break-up events as is done in many studies might be still experiencing the effect of the initial conditions.


Beyond this uncertainty of non-physical initial conditions, studies using decaying turbulence cannot separate the effect of the non-equilibrium change in the global properties of the surrounding carrier phase turbulence on the extracted statistics of the fragmentation processes for droplets/bubbles with timescales commensurate with some of the energetic scales of turbulence. In particular, for bubbles/droplets placed in the inertial range of turbulence, the recent evidence for the importance of a range of scales of smaller eddies on the break-up process could be drastically different if those small scales are decaying in time \citep{qi2022fragmentation}. Controlling this non-equilibrium effects would allow for more accurate measurements of the equilibrium statistics against which the non-equilibrium cases can be compared and interpreted as either slight deviations or completely different phenomena. As such, a statistically stationary state is sought instead. 
To achieve this state, forcing is necessary to sustain the continuously dissipating turbulence, ensuring that the turbulence properties do not change during the breakup processes. 
In the cases where forcing has been used, both simultaneous (both phases)
and individual phase (carrier), linear, physical-space forcing \citep{lundgren2003linearly,rosales2005linear} have been utilized in various studies. However, individual phase (carrier) forcing is only used for the cases that study the break-up of single bubbles in HIT \citep{riviere2021sub,riviere2022capillary}, and are hence more likely influenced by the choice of the initial condition as the simulations are only performed up to first few break-up events.
In contrast, simultaneous mixture-phase (both phases) forcing has been applied only to density ratios representative of emulsions \citep{boukharfane2021structure,begemann2022effect,crialesi2022modulation,krzeczek2023effect,crialesi2023interaction,boniou2025revisiting}. As such, there remain open questions concerning the differing effects of simultaneous mixture-phase (both phases) forcing and individual phase (carrier) forcing, as well as how the aggregate properties of the two-phase HIT vary with density and viscosity ratios when forced.  

In this work, we perform forced numerical simulations of two-phase HIT that will primarily belong to the second category in the classification of existing studies above. We formulate various forcing schemes, including mixture-phase and individual-carrier-phase forcing for two-phase turbulent flows combined with proportional controllers \citep{bassenne2016constant}, and use them to characterize the aggregate statistical properties and the phenomenological behavior of the mixture as a function of different parameters. The parameter space spanned includes suspended heavier drops in turbulence, lighter bubbles in turbulence, and turbulent emulsions. 
In this study, we aim to specifically examine these fundamental issues: 
\begin{itemize}
    \item How are the interfacial statistics, e.g., surface area, influenced during the initial few eddy turnover times by the initialization inside the dispersed phase? 
    \item How do global aggregate statistics of the two-phase mixture vary with varying density and viscosity ratios, void fraction, and Weber number, in stationary state? What is the effect of different forcing schemes on these global aggregate statistics? 
    \item How is turbulent kinetic energy distributed at different scales inside the dispersed phase and in the carrier phase? How do the energy spectra in the dispersed and carrier phases vary for dilute and dense suspensions? At varying density and viscosity ratios? 
    \item Do bubbles and drops possess different turbulence characteristics?
\end{itemize}
Analyzing the aggregate statistics of the two-phase turbulent mixture in a stationary state will help understand the global behavior of the two-phase mixture and will influence future choices in designing cases to probe the more detailed dynamics of the two-phase turbulent breakup and coalescence.
These fundamental questions aim to enhance our understanding of interfacial dynamics and its interaction with turbulence, thereby informing the design and optimization of various industrial applications and natural processes. 

The rest of the paper is structured as follows: Section \ref{methodology} presents the governing equations and numerical methods, including the phase-field method (Section \ref{sec:phase-field}) used, the consistent momentum equations (Section \ref{sec:momentum}), and the surface tension model (Section \ref{sec:surface-tension}). 
Section \ref{sec:energetics} summarizes the mechanisms of energy transfer in incompressible two-phase flows, and presents the two-phase mean mixture energy equations (Section \ref{sec:mixture-energy}) and phase-averaged energy equations (Section \ref{sec:individual-energy}). 
Section \ref{sec:force} presents formulations of the various forcing schemes.
Section \ref{sec:spectrum} presents the spatial filtering approach to calculate the dispersed and carrier phase spectra in two-phase turbulent flows.
Next, the computational setup (Section \ref{sec:setup}), parameters studied (Section \ref{sec:parameters}), and the effect of different forcing schemes on the stationary state of the two-phase turbulent mixture (Section \ref{sec:forcing_effect}) are presented in detail in Section \ref{sec:numerical-experiments}. This includes the discussion on the effect of the initial condition inside the dispersed phase on the global quantities, and the timescales over which this dependence subsists (Section \ref{sec:initialization}). Section \ref{sec:results} presents the simulation results and the corresponding analyses, including the discussions on the variation of the global quantities as non-dimensional parameters such as void fraction, density and viscosity ratios, and large-scale Weber number are varied. Furthermore, the variation of dispersed and carrier phase spectra as a function af these parameters and the size distribution as a function of large-scale Weber number are also discussed. Finally, a summary of the findings and concluding remarks are presented in Section \ref{conclusions}.

\section{Governing system of equations for two-phase turbulent flows}\label{methodology}


In this work, we consider incompressible two-fluid flows. A phase-field/diffuse-interface method is used as an interface-capturing method. The details of the phase-field formulation, the consistent momentum equation, and the improved surface-tension model, used in this work, are presented in Sections \ref{sec:phase-field}-\ref{sec:surface-tension}.

\subsection{Phase-field model \label{sec:phase-field}}

In this method, the volume fraction of one of the fluids is indicated by the phase-field variable $\phi=\phi_1$, which satisfies the relation $\sum_{l=1,2}\phi_l$=1 by definition, where the subscript $l$ denotes phase index. Then, the accurate conservative diffuse-interface (ACDI) model for $\phi$ \citep{jain2022accurate} can be written as
\begin{equation}
\frac{\partial \phi}{\partial t} + \frac{\partial \phi u_j}{\partial x_j} = \frac{\partial }{\partial x_j} \left\{\Gamma\left\{\epsilon \frac{\partial \phi}{\partial x_j}  - \frac{1}{4} \left[1 - \tanh^2{\left(\frac{\psi}{2\epsilon}\right)}\right]\frac{\partial \psi/\partial x_j}{|\partial \psi/\partial x_j|}\right\}\right\},
\label{eq:acdi}
\end{equation}
where $u_i$ is the velocity, $i$ and $j$ are the Einstein indices, $\Gamma$ denotes the velocity-scale parameter, and $\epsilon$ is the interface thickness-scale parameter. 
Here, $\psi$ is an auxiliary variable and represents the signed-distance-like function from the interface, which is defined as
\begin{equation}
    \psi = \epsilon \ln\left(\frac{\phi + \varepsilon}{1 - \phi + \varepsilon}\right).
    \label{eq:psi}
\end{equation}
Note that a small number $\varepsilon$ is added to both the numerator and denominator to 
avoid division by 0, and $\varepsilon$ is chosen as $\varepsilon = 10^{-100}$.
The right-hand-side (RHS) term of Eq.~\eqref{eq:acdi} is the interface-regularization term, which is an artificial term that is added to maintain the constant thickness of the interface. The interface-regularization term contains a diffusion and a sharpening term, which balance each other to maintain a constant interface thickness on the order of $\epsilon$. The ACDI model is discretized using a second-order skew-symmetric split central scheme \citep{jain2022kinetic} that has superior numerical properties, such as improved robustness and lower aliasing error.

\cite{jain2022accurate} showed that $\phi$ remains bounded between $0$ and $1$ using a central scheme, provided $\Gamma/|\vec{u}|_{max}\ge 1$ and $\epsilon/\Delta > 0.5$ conditions are met, where $\Delta$ represents the grid size. Therefore, in this work, the values of $\Gamma/|\vec{u}|_{max} = 1$ and $\epsilon/\Delta = 0.51$ are used for all the simulations, which result in maintaining the bounds of $\phi$.

This recently developed ACDI method has been shown to be suitable for complex turbulent flow simulations \citep{jain2022accurate,hwang2024robust} and a wide range of other multiphysics flow problems \citep{Collis2022, scapin2022mass, brown2023phase, jain_solsurf2023, jain2024model}. It is known to be more accurate than other existing phase-field models because it maintains a sharper interface\textemdash with only one-to-two grid points across the interface\textemdash while being non-dissipative, robust, less expensive, and conservative, without the need for any geometric treatment. Because of these advantages, the ACDI method is chosen as the interface-capturing method in this work.

\subsection{Momentum equation \label{sec:momentum}}
A consistent and conservative momentum equation is solved for the ACDI method that results in consistent momentum transport with the phase-field variable. This includes the effect of the interface-regularization term on the RHS of Eq. \eqref{eq:acdi} in the momentum equation as
\begin{equation}
\frac{\partial \rho u_i}{\partial t} + \frac{\partial \rho u_i u_j}{\partial x_j} =  -\frac{\partial p}{\partial x_i} + \frac{\partial \tau_{ij}}{\partial x_j} + \frac{\partial u_i f_j}{\partial x_j} + \sigma \kappa \frac{\partial \phi}{\partial x_i},  
\label{eq:momf}
\end{equation}
where $\tau_{ij}$ is the stress tensor expressed as $\tau_{ij} = 2\mu S_{ij} = \mu(\partial u_i / \partial x_j+\partial u_j / \partial x_i)$, $p$ is the pressure, and $f_i$ is the implied artificial mass flux that can be written as
\begin{equation}
    f_i= \left\{ (\rho_1-\rho_2) \Gamma \left\{\epsilon \frac{\partial \phi}{\partial x_i}  - \frac{1}{4} \left[1 - \tanh^2{\left(\frac{\psi}{2\epsilon}\right)}\right] n_i \right\} \right\},
    \label{eq:artificial mass flux}
\end{equation}
where $n_i=(\partial \psi/\partial x_i)/\sqrt{(\partial \psi/\partial x_j) (\partial \psi/\partial x_j})$ is the interfacial normal vector.
The Eq. \eqref{eq:momf} is solved along with the continuity equation, $\partial u_j/\partial x_j=0$, that maintains the divergence-free condition for incompressible two-phase flows. This system, when used with the ACDI model, is shown to be robust for the simulations of multiphase turbulent flows, even in the limit of infinite Reynolds numbers, due to the conservation of global kinetic energy \citep{jain2022accurate}. 

\subsection{Surface tension model \label{sec:surface-tension}}

The surface tension forces, $f^\sigma_i = \sigma \kappa (\partial \phi/ \partial x_i)$ [last term in Eq. \eqref{eq:momf}], are modeled using a continuum-surface force (CSF) formulation \citep{Brackbill1992}, where
$\sigma$ is the surface tension coefficient and $\kappa$ is the surface curvature, defined as 
\begin{equation*}
    \kappa = -(\partial n_j/\partial x_j) =-\partial/\partial x_j \left[(\partial \psi/\partial x_j)/\sqrt{(\partial \psi/\partial x_i) (\partial \psi/\partial x_i}) \right].
\end{equation*}
The use of $\psi$, as opposed to $\phi$, to compute curvature in the surface tension force was shown in \citet{jain2022accurate} to improve modeling of the surface tension force and reduce spurious currents significantly. This \textit{improved CSF} model results in lower spurious currents compared to other existing surface tension models for diffuse-interface methods in the literature, including the free-energy based surface tension models \citep{jain2023implementation}. Hence, the \textit{improved CSF} formulation is used to model surface tension forces in this work.

\section{Two-phase energetics \label{sec:energetics}}

In this section, we first derive the incompressible two-phase energy equations to illustrate the different ``buckets" of energy in two-phase flows and the energy transfer mechanisms between them. The internal and kinetic energies in the two phases can be combined together as mixture internal energy and mixture kinetic energy, to study the energy exchange between them and with the surface energy. The equations for these quantities are derived in Section \ref{sec:mixture-energy}. On the other hand, the internal and kinetic energies of the two phases can be kept separate to study the interactions between the kinetic and internal energies of each phase, as well as with the surface energy. The equations for these quantities are derived in Section \ref{sec:individual-energy}.

\subsection{Mixture energy equations \label{sec:mixture-energy}}


Starting from the momentum equation in Eq. \eqref{eq:momf}, taking the dot product with $u_i$, we can arrive at the turbulent kinetic energy (TKE) equation for the mixture (both phases combined) as 
\begin{equation}
\begin{gathered}
\frac{\partial}{\partial t}\left(\rho \frac{u_i u_i}{2}\right)+\frac{\partial}{\partial x_j}\left(u_j \rho \frac{u_i u_i}{2}\right)=- \frac{\partial u_j p}{\partial x_j} 
+u_i f^\sigma_i + \frac{\partial}{\partial x_j}\left(f_j \frac{u_i u_i}{2}\right) + \frac{\partial u_i \tau_{i j}}{\partial x_j} - \tau_{i j} \frac{\partial u_i }{\partial x_j},
\end{gathered}
\label{eq:TKE}
\end{equation}
where $u_i$ represents the fluctuating velocity ($u'$ - primes are dropped everywhere in this paper because the mean velocity is zero). Averaging Eq. \eqref{eq:TKE} in space, we get the mean mixture TKE equation as
\begin{equation}
    \frac{\partial k}{\partial t} = \Psi - \epsilon,
    \label{eq:mean_TKE}
\end{equation}
where $k=\langle \rho u_i u_i/2 \rangle$ is the mean mixture TKE, $\Psi = \langle u_i f_i^{\sigma} \rangle$ is the mean surface tension power, and $\epsilon = \langle \tau_{ij} {\partial u_i }/{\partial x_j} \rangle$ is the dissipation; for all these terms, the averaging operator is defined as $\langle \cdot \cdot \rangle = \left[\int_\Omega (\cdot \cdot) d\Omega\right]/\Omega$, where $\Omega$ is the domain volume. Similarly, we can write the mean surface energy equation \citep{dodd2016interaction} as
\begin{equation}
    \frac{\partial se}{\partial t} = - \Psi,
    \label{eq:mean_SE}
\end{equation} 
where $se = \langle \sigma \delta \rangle$ is the mean surface energy (SE), for which $\delta$ is the local three-dimensional Dirac-delta function active at the interface location \citep{Tryggvason2011}. If we define mean total energy as the sum of TKE and SE, then summing up Eqs. \eqref{eq:mean_TKE} and \eqref{eq:mean_SE}, we get the total energy equation as
\begin{equation}
    \frac{\partial k}{\partial t} + \frac{\partial se}{\partial t} = - \epsilon.
    \label{eq:mean_total_energy}
\end{equation}

\begin{figure}
    \centering
    \includegraphics[width=0.6\textwidth]{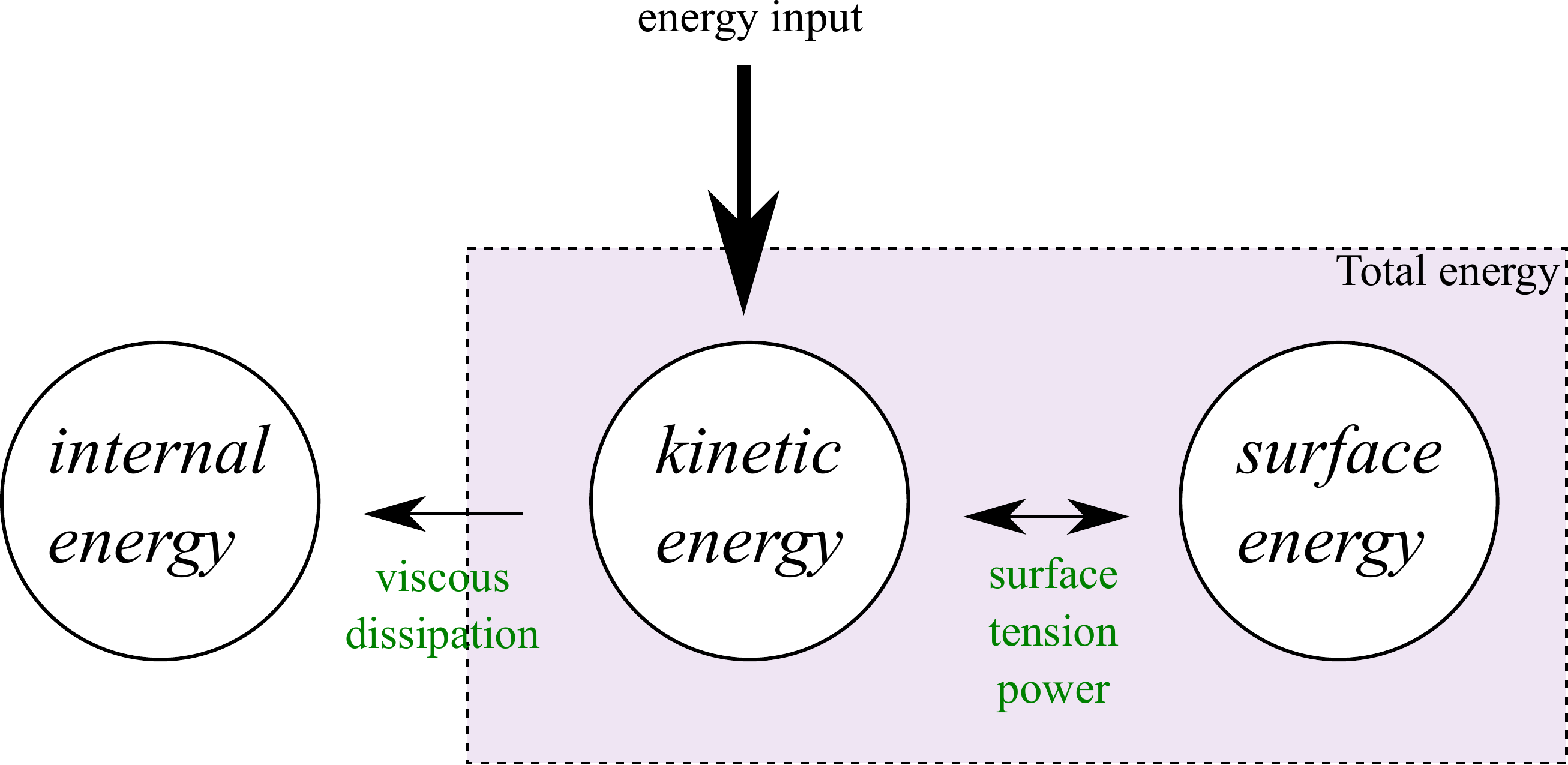}
    \caption{A schematic representing the different energy buckets in an incompressible two-phase flow and the mechanisms responsible for the exchange between them. Here, the kinetic and internal energies represent the mixture quantities, where the energy for both phases is combined. The total energy in the shaded region represents the sum of kinetic and surface energies. The energy input represents the energy injection point corresponding to the external forcing.}
    \label{fig:MixedEnergyBuckets} 
\end{figure}

In Eq. \eqref{eq:mean_TKE}, $\Psi$ is responsible for energy exchange between TKE and SE and can act as a sink or source, and therefore, the same term with an equal and opposite sign is present in Eq. \eqref{eq:mean_SE}. The $\epsilon$ term is the dissipation that acts as a sink and takes away energy from TKE. Figure \ref{fig:MixedEnergyBuckets} is a schematic illustrating the energy buckets for an incompressible two-phase flow mixture, and the energy transfer mechanisms between the buckets.

\subsection{Phase-averaged energy equations\label{sec:individual-energy}}


To derive the energy equations separately for each of the phases, we start with the individual phase momentum equations using a two-fluid formulation as
\begin{equation}
    \frac{\partial \rho_l u_{i,l}}{\partial t} + \frac{\partial \rho_l u_{i,l} u_{j,l}}{\partial x_j} = -\frac{\partial p_l}{\partial x_i} + \frac{\partial \tau_{ij,l}}{\partial x_j}.
    \label{eq:mom_phase}
\end{equation}
The jump condition for the stresses across the interface can be written as
\begin{equation}
    \sum_l \left(  -p_l \delta_{ij} + \tau_{ij,l} \right) n_{j,l} = \sigma \kappa n_{i,l},
    \label{eq:jump_condition}
\end{equation}
which indicates that the net normal viscous and pressure forces at the interface balance the surface tension forces. Then, the TKE equation for each phase $l$ can be obtained by taking the dot product of Eq. \eqref{eq:mom_phase} with $u_{i,l}$ as
\begin{equation}
\begin{gathered}
\frac{\partial}{\partial t}\left(\rho_l \frac{u_{i,l} u_{i,l}}{2}\right)+\frac{\partial}{\partial x_j}\left(u_{j,l} \rho_l \frac{u_{i,l} u_{i,l}}{2}\right)+\frac{\partial u_{j,l}  p_l}{\partial x_j} = \frac{\partial u_{i,l} \tau_{i j,l}}{\partial x_j} - \tau_{i j,l} \frac{\partial u_{i,l} }{\partial x_j}.
\end{gathered}
\label{eq:TKE_phase}
\end{equation}
If we now define volume averaging within each phase $l$ as
\begin{equation*}
    \langle \cdot \cdot \rangle_l = \frac{1}{\Omega_l} \int_{\Omega_l} (\cdot \cdot) d\Omega_l = \frac{1}{\Omega_l} \int_{\Omega} (\phi_l \cdot \cdot) d\Omega = \frac{\Omega}{\Omega_l} \langle \phi_l \cdot \cdot \rangle = \frac{\langle \phi_l \cdot \cdot \rangle}{\langle \phi_l \rangle},
\end{equation*}
where $\cdot \cdot$ represents any generic quantity, $\Omega_l$ is the phase volume, and $\phi_l$ is the volume fraction of phase $l$; and average Eq. \eqref{eq:TKE_phase} in space within each phase $l$, we get the phase-averaged TKE equations as
\begin{equation}
    \frac{\partial k_l}{\partial t} = V_l - P_l - \epsilon_l,
    \label{eq:mean_phase_TKE}
\end{equation}
where $k_l=\langle \phi_l \rho_l u_{i,l} u_{i,l}/2 \rangle/\langle \phi_l \rangle$ is the phase-averaged TKE, 
$V_l = \langle \phi_l \left( {\partial u_{i,l} \tau_{i j,l}}/{\partial x_j} \right) \rangle/\langle \phi_l \rangle$ is the phase-averaged viscous transport contribution, 
$P_l = \langle \phi_l \left( {\partial u_{j,l} p_l}/{\partial x_j} \right) \rangle/\langle \phi_l \rangle$ is the phase-averaged pressure transport contribution, and 
$\epsilon_l = \langle \phi_l \tau_{ij} \left( {\partial u_{i,l} }/{\partial x_j} \right) \rangle /\langle \phi_l \rangle$ is the dissipation of phase $l$. These equations are consistent with the diffuse-interface/phase-field formulation, where $\phi_l$ here represents the phase-field variable [see \citet{dodd2016interaction} for a sharp-interface formulation of phase-averaged TKE equations].

\begin{figure}
    \centering
    \includegraphics[width=0.6\textwidth]{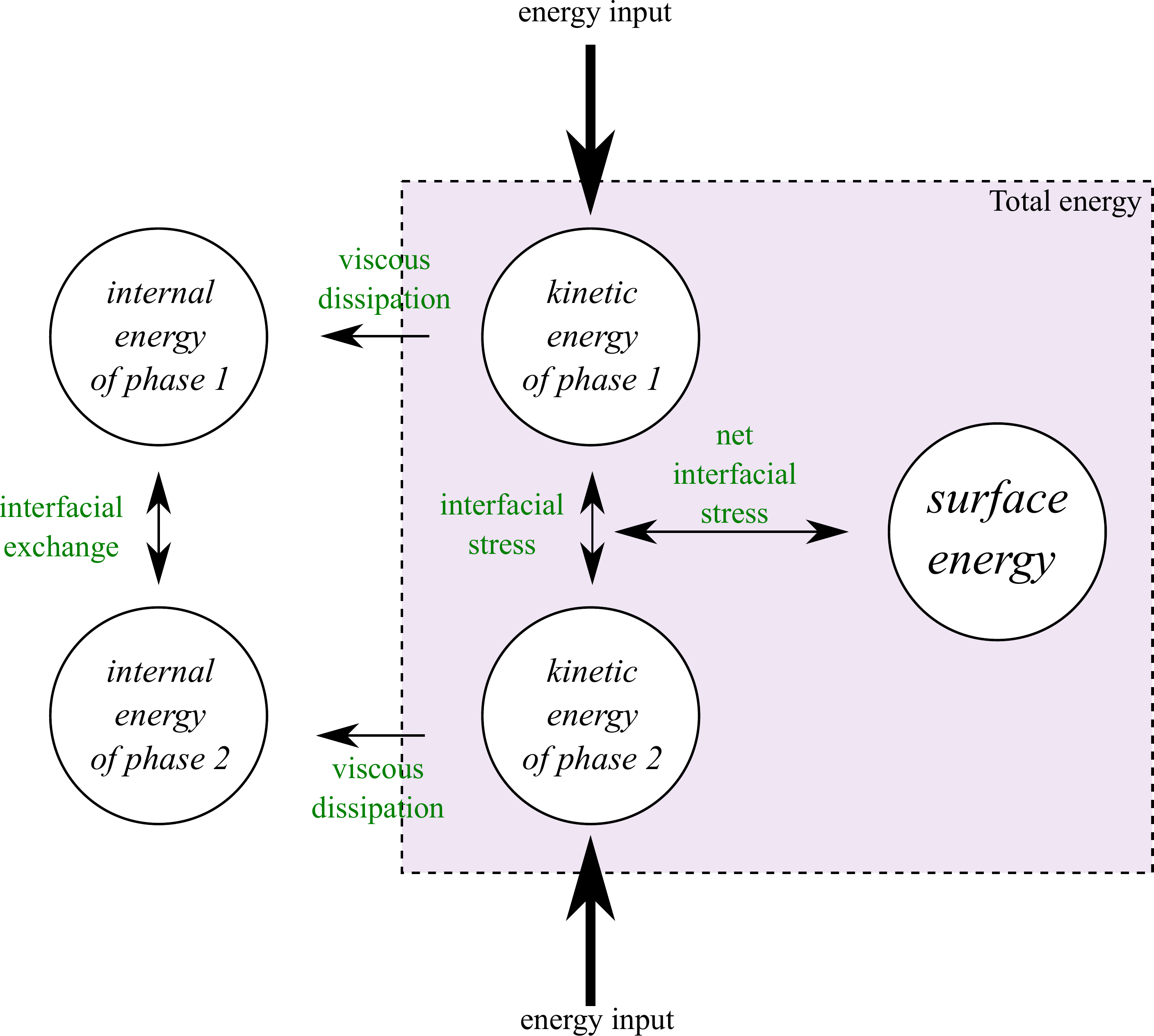}
    \caption{A schematic representing the different energy buckets in an incompressible two-phase flow (with the individual phases represented separately) and the exchange terms between them. The total energy in the shaded region is a sum of the kinetic energies of all phases and the surface energy. The energy injection points correspond to the external forcing to each of the phases.}
    \label{fig:PhaseEnergyBuckets}
\end{figure}

Figure \ref{fig:PhaseEnergyBuckets} is a schematic illustrating the individual-phase energy buckets available in an incompressible two-phase flow, and the energy transfer mechanisms between the buckets. There is an exchange of TKE between phase 1 and phase 2 through the interfacial stresses, which are normal viscous and pressure stresses [$V_l$ and $P_l$ in Eq. \eqref{eq:mean_phase_TKE}], and the net interfacial stress [Eq. \eqref{eq:jump_condition}] contributes toward the surface energy and vice versa. The dissipation $\epsilon_l$ in Eq. \eqref{eq:mean_phase_TKE} is a sink term responsible for transferring energy from TKE to the internal energy of each phase.


\section{Forcing formulations for two-phase turbulence \label{sec:force}}


To maintain the turbulent state of the two-phase flow, we force the system. Broadly, there are two possible ways to force a two-phase system: (a) forcing is active simultaneously in both phases (mixture forcing), and (b) forcing is active in one of the phases (individual phase forcing). 

\subsection{Mixture and individual phase forcings \label{sec:forcing_approaches}}
Both phases in the two-phase system can be forced by adding a forcing source term $F_i$ to the momentum equation in Eq. \eqref{eq:momf} as 
\begin{equation}
\frac{\partial \rho u_i}{\partial t} + \frac{\partial \rho u_i u_j}{\partial x_j} =  -\frac{\partial p}{\partial x_i} + \frac{\partial \tau_{ij}}{\partial x_j} + \frac{\partial u_i f_j}{\partial x_j} + \sigma \kappa \frac{\partial \phi}{\partial x_i} + F_i.  
\label{eq:momf_force}
\end{equation}
and the resulting mean mixture TKE equation in Eq. \eqref{eq:mean_TKE} with the forcing will be
\begin{equation}
    \frac{\partial k}{\partial t} = \Psi - \epsilon + \langle F_i u_i \rangle,
    \label{eq:mean_TKE_F}
\end{equation}
and the mean total energy equation can be obtained by summing Eq. \eqref{eq:mean_TKE_F} and Eq. \eqref{eq:mean_SE} as
\begin{equation}
    \frac{\partial k}{\partial t} + \frac{\partial se}{\partial t} = - \epsilon + \langle F_i u_i \rangle.
    \label{eq:mean_TE_F}
\end{equation}
Note that this forcing in Eq. \eqref{eq:momf_force} only modifies the mean mixture TKE equation and will not directly modify the exchange between kinetic and surface energy through surface tension power, i.e., Eq. \eqref{eq:mean_SE} is unaltered. So, the dynamics of the multiphase system, such as breakup and coalescence, will not be directly altered due to the forcing. 
One can also potentially design an alternative forcing approach (that is active only at the interface) to modify the mean surface energy in the system directly. But this may modify the energy transfer at the interface, and alter the breakup and coalescence of drops/bubbles. This could be unphysical, and hence, this approach is not explored in this work. 

Forcing an individual phase in the two-phase system can be achieved by including a source term $F_{i,l}$ in the individual phase momentum equation in Eq. \eqref{eq:mom_phase} as
\begin{equation}
    \frac{\partial \rho_l u_{i,l}}{\partial t} + \frac{\partial \rho_l u_{i,l} u_{j,l}}{\partial x_j} = -\frac{\partial p_l}{\partial x_i} + \frac{\partial \tau_{ij,l}}{\partial x_j} + F_{i,l},
    \label{eq:mom_phase_force}
\end{equation}
and the resulting phase-averaged TKE equation in Eq. \eqref{eq:mean_phase_TKE} with forcing will be
\begin{equation}
    \frac{\partial k_l}{\partial t} = V_l - P_l - \epsilon_l + \frac{\langle \phi_l F_{i,l} u_{i,l} \rangle}{\langle \phi_l \rangle}.
    \label{eq:mean_phase_TKE_F}
\end{equation}

\subsection{Constant-energetics physical space linear forcing}
While there are many possible forms for the forcing terms $F_i$ and $F_{i,l}$, it has been shown that a linear forcing term in momentum applied in physical space generates more favorable statistics of third-order structure functions, and presumably other inertial range statistics, as compared with low-wavenumber forcing \citep{lundgren2003linearly}. Given that we wish to study the breakup and coalescence of droplets in the inertial range of turbulence, this fact, combined with the ease of implementation, leads us to use linear forcing. If we adopt linear forcing of the form $F_i = \rho A u_i$ for the mixture momentum equation and $F_{i,l} = \rho_l A u_{i,l}$ for the phase momentum equation, where $A$ is a constant with the units of time, then the forcing terms in the energy equations Eq. \eqref{eq:mean_TKE_F} and Eq. \eqref{eq:mean_phase_TKE_F} will be 
\begin{equation}
    \langle F_i u_i \rangle \equiv A \langle \rho u_i u_i \rangle = 2Ak\hspace{0.5cm} \mathrm{and} \hspace{0.5cm}
    \frac{\langle \phi_l F_{i,l} u_{i,l} \rangle}{\langle \phi_l \rangle} \equiv A \frac{\langle \phi_l \rho_l u_{i,l} u_{i,l} \rangle}{\langle \phi_l \rangle} = 2Ak_l.
    \label{eq:force_A}
\end{equation}

We also use a forcing controller \citep{bassenne2016constant} to efficiently achieve a constant energy state of the system. If $dq/dt=T + F$ is a generic dynamical system, where $T$ is a physical term and $F$ represents the forcing term, then the force that maintains a constant stationary value of $q$ can be obtained by setting $dq/dt=0$, which results in $F=-T$. Now, if we seek to achieve the stationary state exponentially, then we need $dq/dt=-G(q-q_{\infty})/t_{\infty}$ instead, which can be achieved by adding a controller to this forcing term as
\begin{equation}
    F = - T - G\left(\frac{q - q_{\infty}}{t_{\infty}} \right).
\end{equation}
where $G$ is a constant that controls the strength of the controller, $q_{\infty}$ is the stationary state value of $q$, and $t_{\infty}$ is the relaxation timescale for the system. 

In a two-phase system, there are three choices for the quantity that can be held constant when the system reaches a stationary state: mean mixture TKE, mean total energy, and individual phase-averaged TKE. The details of the forcing formulation to be used to maintain a constant value for these three quantities are described below.

\textbf{Constant TKE forcing:} To maintain a constant value of the mean mixture TKE in the system, the right-hand-side (RHS) terms in Eq. \eqref{eq:mean_TKE_F} has to go to zero. So if we set, $\langle F_i u_i \rangle = \epsilon - \Psi$, then the value of the constant $A$ in the momentum forcing term, can be calculated using Eq. \eqref{eq:force_A}, and is given by  
\begin{equation}
    A = \frac{\epsilon - \Psi}{2k},
\end{equation}
or with the use of the controller,
\begin{equation}\label{Const_TKE_Forcing}
    A = \frac{\epsilon - \Psi - G\left(\frac{k - k_{\infty}}{t_{\infty}}\right)}{2k}.
\end{equation}

\textbf{Constant total energy forcing:} To maintain a constant value of the mean mixture total energy (TKE + SE) in the system, the RHS terms in Eq. \eqref{eq:mean_TE_F} has to go to zero. So if we set, $\langle F_i u_i \rangle = \epsilon$, then the value of the constant $A$ in the momentum forcing term, can be calculated using Eq. \eqref{eq:force_A}, and is given by 
\begin{equation}
    A = \frac{\epsilon}{2k},
\end{equation}
or with the use of the controller,
\begin{equation}\label{Const_TE_Forcing}
    A = \frac{\epsilon - G\left(\frac{k + se - k_{\infty} - se_{\infty}}{t_{\infty}}\right)}{2k}.
\end{equation}

\textbf{Constant individual phase TKE forcing:} To maintain a constant value of the individual phase-averaged TKE, the RHS terms in Eq. \eqref{eq:mean_phase_TKE_F} has to go to zero. So if we set, ${\langle \phi_l F_{i,l} u_{i,l} \rangle}/{\langle \phi_l \rangle} = \epsilon_l + P_l - V_l$, then the value of the constant $A$ in the momentum forcing term, can be calculated using Eq. \eqref{eq:force_A}, and is given by 
\begin{equation}
    A = \frac{\epsilon_l + P_l - V_l}{2k_l},
\end{equation}
or with the use of the controller,
\begin{equation}\label{Const_TKEP_Forcing}
    A = \frac{\epsilon_l + P_l - V_l - G\left(\frac{k_l - k_{l, \infty}}{t_{\infty}}\right)}{2k_l}.
\end{equation}

The effects of the constant TKE, total energy, and individual-phase TKE forcings on the macroscopic two-phase system behavior are explored in detail in Section \ref{sec:forcing_effect} using numerical simulations. Note that one can also force the two-phase system differently such that a constant surface energy value is maintained in the stationary state. But this would modify the dynamics of the multiphase mixture, as described in Section \ref{sec:forcing_approaches}. Hence, this type of forcing is not explored in this work. Figures \ref{fig:MixedEnergyBuckets}, \ref{fig:PhaseEnergyBuckets} show the forcing inputs for the mixture and for the individual phases in a two-phase system, respectively, as energy injection points in the schematic.

\section{Spatial filtering approach for two-phase spectra \label{sec:spectrum}}

To study the energy content at different scales of each of the dispersed and carrier phases in a two-phase flow, we compute the 
phase-dependent TKE spectra utilizing the spatial filtering approach of \citet{sadek2018extracting}. Note that a standard Fourier spectrum is not a general enough approach for when there are variations in density and viscosity between the two phases. This is due to three reasons: First, the viscosity variations introduce discontinuities of velocity derivatives at interfaces, contaminating the spectrum. Second, using a Fourier spectrum, one cannot isolate the dispersed phase and carrier phase contributions. Third, and perhaps most importantly, energy is a cubic quantity in compressible flows and volumetrically filtered variable-density flows such as non-unity density ratio incompressible two-phase flows. However, energy must be defined quadratically using a Fourier spectrum; no longer satisfying the inviscid criterion \citep{aluie2013scale,zhao2018inviscid}, i.e., the lack of influence of viscous effects at the large scales. 

Using the wavelet method of \citet{freund2019wavelet}, spectra have been computed before for the dispersed and carrier phases for some wavenumbers in an incompressible two-phase flow. However, due to the need for decomposing spatial regions into carrier, dispersed, and interaction phases, the range of scales over which a dispersed phase spectrum can be measured is limited to scales smaller than the average size of the dispersed phase. Moreover, the wavelet-based approach would also not generally work for compressible flows, because similar to a standard Fourier spectrum, the orthogonal wavelet spectrum does not satisfy the inviscid criterion.  However, the spatial filtering approach we propose to compute spectra for two-phase flows in this work generally applies to incompressible/compressible two-phase flows and remedies the issue of the limit on the range of scales examined, as it seamlessly transitions from treating the dispersed phase as a union of disjoint regions to a full Eulerian field.

The spatial filtering approach for a two-phase flow can be applied as follows, inspired by Favre filtering. Let us define the resolved kinetic energy in a variable density field up to a scale $\ell$ as 
\begin{equation}
    \mathcal{E}^{\ell} = \frac{1}{2}\frac{\big(\overline{\rho u_i}^\ell\big)^2}{\overline{\rho}^\ell},
\end{equation}
where $\overline{.}^\ell$ denotes low-pass filtered quantities at scale $\ell$. The energy spectrum can be understood as energy density per unit scale, which is equal to the derivative of the mean $\mathcal{E}^\ell$ with respect to the wavenumber, $k_\ell = 2\pi/\ell$, i.e.,
\begin{equation}
    E^\ell = \frac{d\langle\mathcal{E}^\ell\rangle}{dk_\ell}.
\end{equation}
Since $\rho = \rho_1\phi+\rho_2(1-\phi) = \rho_1\phi_1+\rho_2\phi_2$, we can show that
\begin{multline}
    \mathcal{E}^{\ell} = \frac{1}{2}\bigg(\frac{\big(\overline{\rho_1\phi_1 u_i}^\ell\big)^2}{\overline{\rho_1\phi_1}^\ell} + \frac{\big(\overline{\rho_2\phi_2 u_i}^\ell\big)^2}{\overline{\rho_2\phi_2}^\ell} - \frac{\overline{\rho_1\phi_1}^\ell\overline{\rho_2\phi_2}^\ell}{\overline{\rho_1\phi_1}^\ell+\overline{\rho_2\phi_2}^\ell}\big(\tilde{u^1_i}^\ell-\tilde{u^2_i}^\ell\big)^2\bigg)
    \\ = \mathcal{E}^\ell_{\phi_1}+\mathcal{E}^\ell_{\phi_2}-\mathcal{E}^\ell_{\phi_{1,2}}
    \label{eq:cum-energy}
\end{multline}
where $\tilde{u}^\ell$ is the Favre-filtered velocity of a particular phase, and where the local energy at some scale is now a sum of three components. The first two are the explicit contributions from each of the individual phases, $\mathcal{E}^\ell_{\phi_\alpha}$. The third contribution, $\mathcal{E}^\ell_{\phi_{\alpha,\beta}}$, is a positive semi-definite quantity and is subtracted from the other two. It represents the energy due to the mixture of the two phases at a point and is only present when there is a relative velocity between the Favre-averaged velocities of the individual phases. 
Taking the gradient of the volumetric mean of the three quantities with respect to the filtering scale provides us with a precise way of defining the phase-dependent and interaction spectra even in the presence of first-order discontinuities due to non-unity density and viscosity ratios without the need for user-defined spatial partitioning of the interaction, and phase-dependent spectra. For the spectra presented in this work, the filtering kernel is chosen to be Gaussian to ensure that the filtered phase field variables remain positive definite. The length scale $\ell$ is defined to be the full-width at half-maximum of the Gaussian kernel.

\section{Numerical experiments with two-phase isotropic turbulence \label{sec:numerical-experiments}}

In this work, we perform forced two-phase homogeneous isotropic turbulence (HIT) simulations to study the various stationary states of the system. A stationary two-phase turbulent state is achieved by embedding a bubble or a drop in a HIT and by forcing the system using one of the formulations described in Section \ref{sec:force}.   

\subsection{Computational setup \label{sec:setup}}

For all the simulations presented in this study, the computational setup consists of a triply periodic domain of size $(2 \pi)^3$ discretized using $N^3$ points with $N = 256$. First, a single-phase flow ($l=1$) is forced using the constant-energetics forcing in Eq. \eqref{Const_TKE_Forcing} with $\Psi = 0$, such that the final Taylor microscale\textendash based Reynolds number, $Re_\lambda$, achieved is $Re_\lambda \approx 87$. This corresponds to a local average resolution of the Kolmogorov length scale, $\eta$, of $k_{max}\eta = 1.5$, where $k_{max}$ is the largest wavenumber resolved by the grid, consistent with the direct numerical simulation of single-phase flow reported in \cite{bassenne2016constant}. This single-phase simulation is run for around $30\tau_e$, where $\tau_e$ is the large eddy turnover time. 

\begin{figure}
    \centering
    \includegraphics[width=0.4\textwidth]{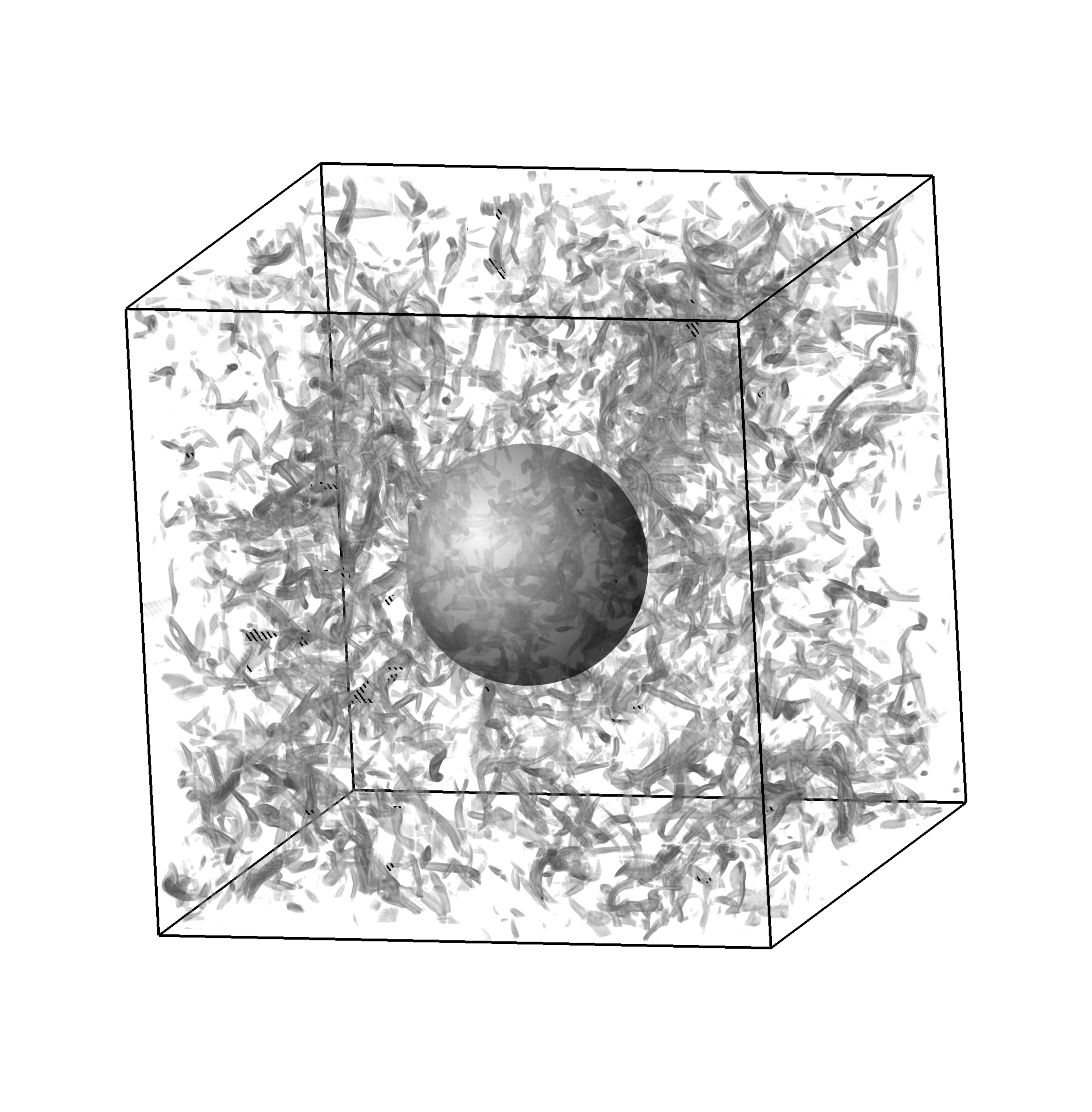}
    \caption{Volume rendering of enstrophy of the HIT along with the interface, showing the initial condition for the two-phase simulations.}
    \label{fig:HIT}
\end{figure}

The final stationary single-phase HIT state at $t\approx30\tau_e$ is used to initialize all the two-phase HIT simulations. This is done by inserting a sphere of another phase, $l=2$ (either a droplet or a bubble) in the center of the domain, as shown in Figure \ref{fig:HIT}. To completely disregard the effect of the initialized velocity field inside the second phase (which could be unphysical), we run this setup using the various controlled forcings until the global surface area of the interface reaches a steady state, after which we collect statistics and make the findings reported below. The grid size of $N=256$ is chosen such that the global surface area is grid converged \citep{boukharfane2021structure,jain2022modeling}. 


\subsubsection{Initialization of the secondary phase: Effect of initial velocity field}\label{sec:initialization}


One approach to initializing a two-phase isotropic turbulence simulation is to first perform a forced single-phase HIT and then add a second phase. When this secondary phase is introduced, the velocity field inside it could be initially unphysical due to the transients associated with its introduction. The flow field will eventually adjust to the new conditions in the presence of two phases. 

Since the velocity field inside the second phase cannot be determined a priori, there have been different approaches to handling this velocity field inside the second phase in the past. \citet{dodd2016interaction} zeroed out the velocity field when droplets of the second phase were introduced into a decaying isotropic turbulence, and \citet{jain2022kinetic} retained the same velocity as in the single-phase HIT when the slab and droplets of the second phase were introduced into an isotropic turbulence simulation.  

\begin{figure}
    \centering
    \includegraphics[width=\linewidth]{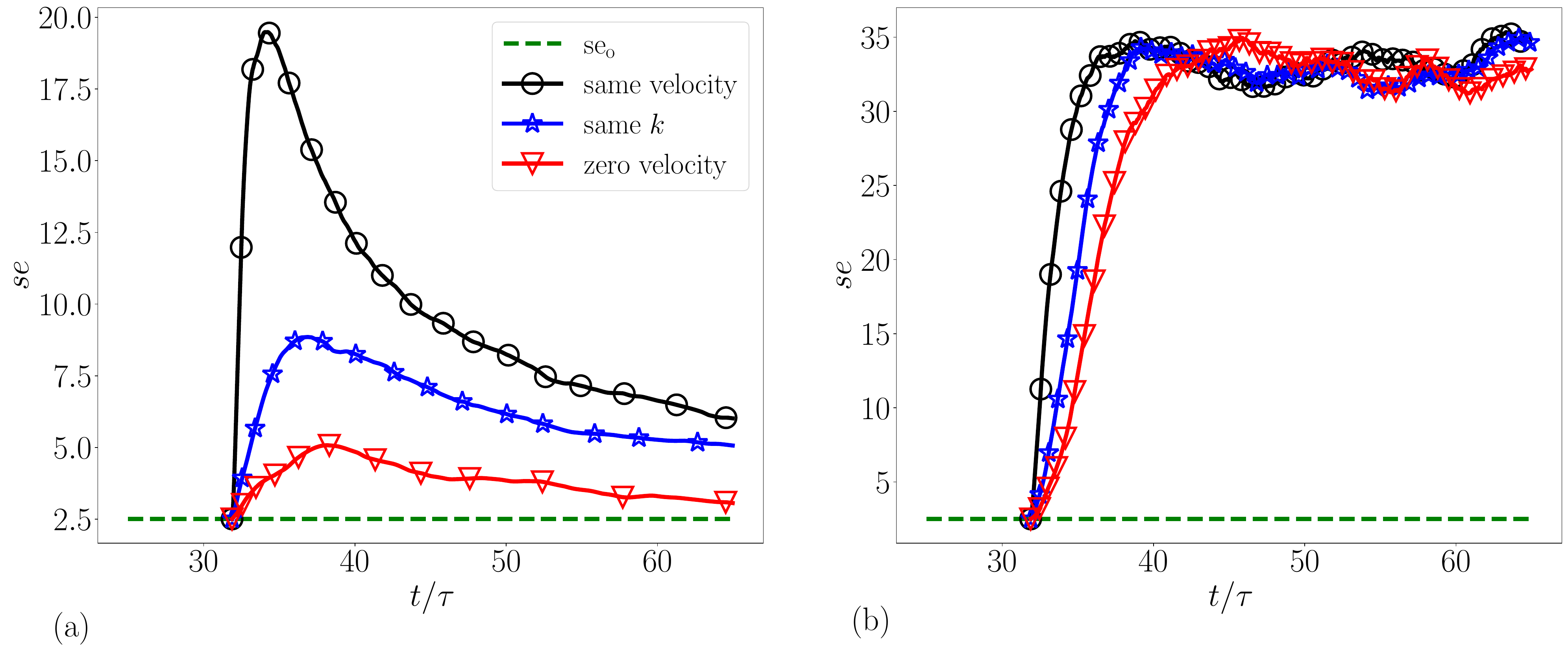}
    \caption{Temporal evolution of the mean surface energy, $se$, with three different approaches to initialize the velocity in the second phase for (a) a decaying simulation without forcing, and (b) a forced simulation.}
    \label{fig:initial-effect}
\end{figure}

When the simulations are not forced, these different approaches to initialize velocity fields could result in different final states of the simulation, which could be arbitrary. Forcing the simulation could help eliminate this arbitrariness. Here, we evaluate the different approaches to initialize the velocity field inside the second phase in a two-phase flow simulation with and without forcing. We explore three different ways: (a) retaining the same velocity for the second phase as in the single-phase case, (b) maintaining the same kinetic energy in the second phase, same as in the single-phase case before the introduction of the second phase, by rescaling the velocity to take into account the change in density, and (c) zeroing out the velocity in the second phase. Figure \ref{fig:initial-effect} shows the effect of these approaches on the temporal evolution of the mean $se$ with and without forcing. Here, a droplet of density $\rho_2=10$ was introduced into the domain consisting of a fluid of density $\rho_1=1$.

As seen in Figure \ref{fig:initial-effect}(a), the flow decays in the absence of forcing, and different approaches to initializing the velocity field result in different temporal evolutions of mean surface energy, and therefore, different final states of the system. Therefore, simulations of decaying two-phase turbulent flows may lead to undesired physical states due to the arbitrariness of the initial condition and its long timescale influence on the system dynamics during which statistics are usually gathered. However, in the presence of forcing, all three approaches reach the same stationary state, but the initial transient behaviors are different [Figure \ref{fig:initial-effect}(b)]. Initializing the second phase with the same velocity as in the single-phase HIT results in reaching the stationary state faster.  
Nevertheless, irrespective of the choice of the forcing used, there appears to be an initial transient evolution of the system that is dependent on the choice of the initial conditions that persists for a few eddy turnover times. Hence, any forced simulation that lasts shorter than this time will again likely be influenced by the choice of the initialization, and therefore, may not be physically accurate.

\subsection{Parameters studied \label{sec:parameters}}

Using the computational setup described here, we run a variety of cases to test the effect of varying non-dimensional parameters such as the density ratio, $\rho_2/\rho_1$, the viscosity ratio, $\mu_2/\mu_1$, the average void fraction of the dispersed phase, $\langle\phi_d\rangle$, and the integral-scale Weber number of the flow defined as
\begin{equation}
    We_L = \left(\frac{2}{3}\right)\frac{\rho_c u_L^2 l_e}{\sigma}.
    \label{eq:We_L}
\end{equation}
Here, $\rho_c$ is the carrier phase density, $l_e$ is the integral length scale of the single-phase flow defined as $l_e=k_e^{3/2}/\epsilon$, where $k_e$ and $\epsilon$ are the targeted kinetic energy and dissipation of the HIT, respectively, and $u_L$ is the root-mean-square (RMS) velocity of the single-phase turbulence before the insertion of the drop/bubble, which is related to the kinetic energy through $k_e = (3/2)u_L^2$. 

Table \ref{tab:TableCases} reports all the properties used for all cases examined in this study. Case A is the base case from which variations in non-dimensional parameters are studied, and all subsequent plots include case A as a reference point along with the corresponding numbered cases for each parameter variation. For case A, the mean volume fraction (void fraction) of phase $2$ of $\langle \phi_2 \rangle=0.0655$ is achieved by initializing a sphere of the second phase of diameter $L_d=\pi$ (Figure \ref{fig:HIT}). For cases C1-C3, $\langle \phi_2 \rangle$ is varied by increasing the diameter of this sphere to $1.5\pi$, $2\pi$, and $3\pi$, respectively.  

For cases B1 and B2, the Weber number values are reduced to $9.75$ and $1.95$, respectively, by increasing $\sigma$ from case A. The Weber number was not further reduced because there was no breakup observed in case B2 [see Figure \ref{fig:we-effect}(c)]. For cases D1-D4, the density of phase 2 was varied while holding the density of phase 1 constant to achieve different density ratios. For cases E1-E4, the viscosity of phase 2 was varied while holding the viscosity of phase 1 constant to achieve different viscosity ratios. In addition, the density ratio was also varied in such a way that the kinematic viscosity is maintained the same in both phases and across cases E1-E4.

\begin{table}
\footnotesize
\centering
\begin{tabular}{ccccccccc}
\toprule
Case & $\rho_2/\rho_1$ &  $\mu_2/\mu_1$ & $\langle \phi_2 \rangle$ & $\langle \phi_d \rangle$ & $We_L$ \\ \midrule
A & $1$ & $1$ & 0.0655 & 0.0655 & 19.5 \\ 
\midrule
\multicolumn{6}{|c|}{Weber number variation} \\ \midrule
B1  & $1$  & $1$ & 0.0655 & 0.0655  & 9.75 \\
B2  & $1$  & $1$ & 0.0655 & 0.0655 & 1.95 \\ 
\midrule
\multicolumn{6}{|c|}{Void fraction variation} \\ \midrule
C1  & $1$  & $1$ & 0.221 & 0.221 &  19.5 \\
C2  & $1$  & $1$ & 0.524 & 0.476 &  19.5 \\
C3  & $1$  & $1$ & 0.738 & 0.262 &  19.5 \\ \midrule
\multicolumn{6}{|c|}{Density ratio variation} \\ \midrule
D1  & $0.01$ &  $1$ & 0.0655 & 0.0655  & 19.5 \\
D2  & $0.1$  & $1$ & 0.0655 & 0.0655  & 19.5 \\
D3 & $10$  & $1$ & 0.0655 & 0.0655  & 19.5 \\
D4 & $100$ & $1$ & 0.0655 & 0.0655 &  19.5 \\ 
\midrule
\multicolumn{6}{|c|}{Viscosity ratio variation} \\ \midrule
E1  & $0.001$  $0.001$ & 0.0655 & 0.0655 &  19.5 \\
E2  & $0.1$ & $0.1$ & 0.0655 & 0.0655 &  19.5 \\
E3 & $10$ & $10$ & 0.0655 & 0.0655 &  19.5 \\
E4 & $1000$  & $1000$ & 0.0655 & 0.0655 &  19.5 \\ 
\bottomrule
\end{tabular}
\caption{\label{tab:TableCases} The properties used for all cases examined in this study: the density of each phase, $\rho_{l}$, along with their ratio, $\rho_2/\rho_1$, the viscosity of each phase, $\mu_{l}$, along with their ratio $\mu_2/\mu_1$, the average volume fraction of phase $2$, $\langle\phi_2\rangle$, along with the dispersed phase volume fraction, $\langle\phi_d\rangle$, the surface tension coefficient, $\sigma$, and the integral-scale Weber number, $We_L$.}
\end{table}



\subsection{Effect of different forcing types \label{sec:forcing_effect}}

Three ways to force a two-phase turbulence system that can hold a constant value of mean mixture TKE, mean total energy (TKE+SE), or the individual phase TKE, when forced, were presented in Section \ref{sec:force}. In this section, we first study the effect of different forcing types on the stationary state of the two-phase turbulence system. 

\begin{figure}
    \centering
    \includegraphics[width=\textwidth]{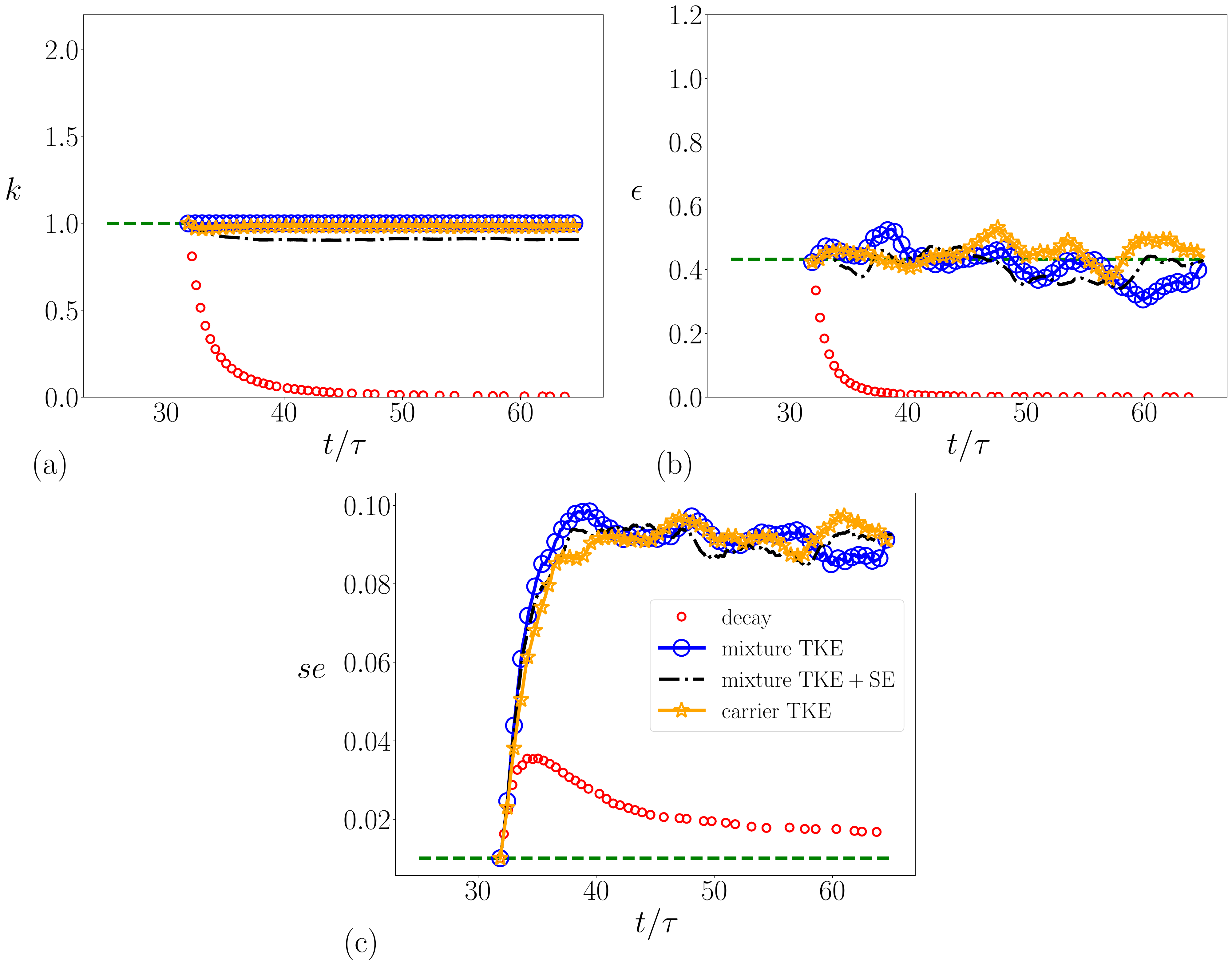}
    \caption{Effect of different forcing schemes on the stationary state of case A, showing (a) mixture TKE, (b) mixture dissipation, and (c) mean surface energy. The legend applies to all the subplots. In the legend, ``decay" refers to a decaying case, ``mixture TKE" refers to a forced case [using the forcing in Eq. \eqref{Const_TKE_Forcing}] with mean mixture TKE held constant, ``mixture TKE+SE" refers to a forced case [using the forcing in Eq. \eqref{Const_TE_Forcing}] with mean total energy held constant, and ``carrier TKE" refers to a forced case [using the forcing in Eq. \eqref{Const_TKEP_Forcing}] with the carrier phase TKE held constant. The dashed reference line is the initial state of the system when the second phase is initialized.}
    \label{fig:DiffForcings}
\end{figure}

For the baseline case A, which has unity-density and unity-viscosity ratios, low void fraction, and a high $We_L$, promoting breakup, we test all three possible forcings, Eqs. \eqref{Const_TKE_Forcing}, \eqref{Const_TE_Forcing}, and \eqref{Const_TKEP_Forcing}, i.e., holding TKE constant, holding TKE + SE constant, and holding only the TKE of the carrier phase constant, respectively, along with the decaying case. Figure \ref{fig:DiffForcings} shows the time evolution of the kinetic energy $k=\langle \rho u_i u_i/2 \rangle$, dissipation $\epsilon = \langle \tau_{ij} {\partial u_i }/{\partial x_j} \rangle$, and the surface energy $se = \langle \sigma \delta \rangle$ of the two-phase mixture after the second phase is introduced in the domain at $t\approx30\tau_e$. 

When the mixture TKE is held constant, both $\epsilon$ and $se$ reach a stationary value. This implies that, on average, the two-phase mixture has reached a steady state, where the breakup and coalescence balance each other, resulting in a steady size distribution of the dispersed phase and a stationary value for the total interfacial area of the two-phase mixture. The energy stored in $se$ is $\approx10\%$ of the energy in the TKE.
When the mixture total energy (TKE+SE) is held constant at a value of $1$, the mixture TKE reaches a stationary value less than $1$, and the remaining energy is stored in $se$. When only the carrier TKE is held constant at a value of $1$, the mixture TKE again reaches a stationary value slightly less than $1$,
which indicates that the kinetic energy in the dispersed phase is less than $1$ even though the density in the carrier and dispersed phases are equal. This is potentially due to the confinement effect of the material interface on the dispersed phase, and will be discussed in more detail in Section \ref{sec:results}. 

Irrespective of the choice of the forcing used, the final value of $se$ is similar, which is due to the small contribution of $se$ to the total energy and is indicative of the potentially negligible effect of $se$ on the kinetic energy of the two-phase mixture.  
Hence, for the remainder of the study, we limit ourselves to constant TKE forcing for either one of the phases or for both phases simultaneously. Forcing one of the phases vs both phases simultaneously can make a difference when the densities of the phases are not the same. This will be explored further in Section \ref{sec:density}.


\begin{figure}
    \centering
    \includegraphics[width=\linewidth]{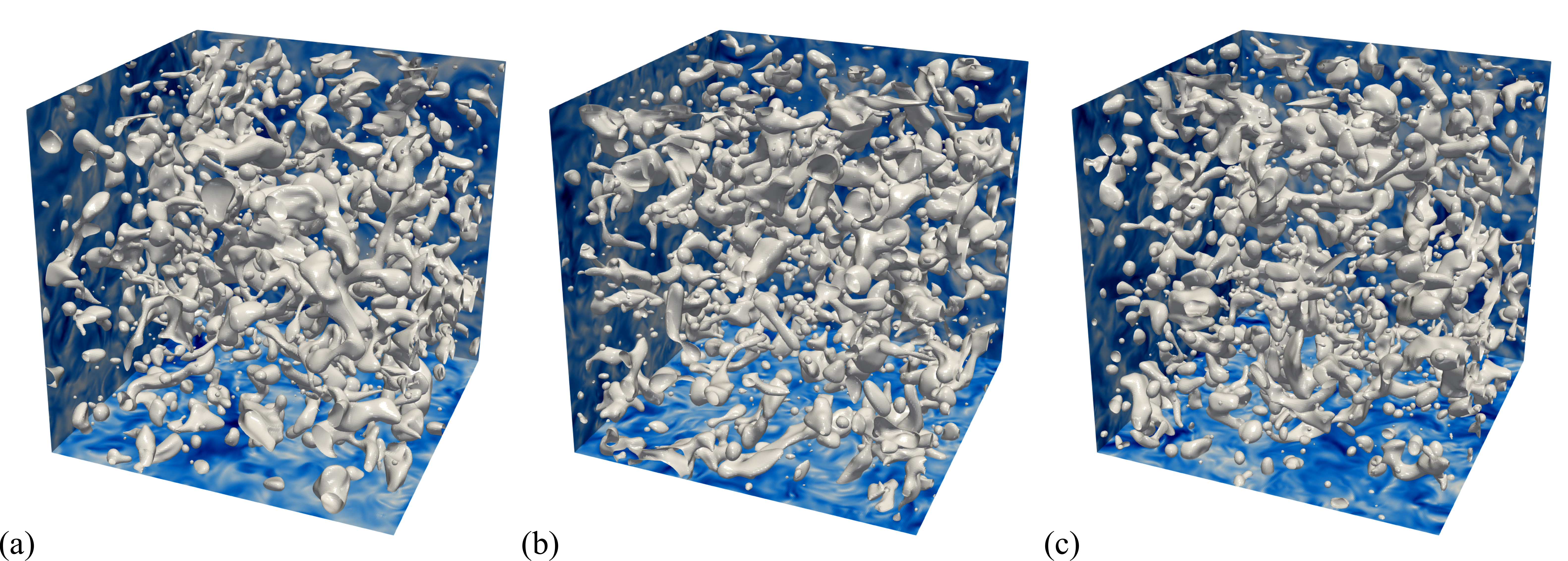}
    \caption{Visualization of the flow fields in stationary state for case A with (a) the forcing in Eq. \eqref{Const_TKE_Forcing} with mean mixture TKE held constant, (b) the forcing in Eq. \eqref{Const_TE_Forcing} with mean total energy held constant, and (c) the forcing in Eq. \eqref{Const_TKEP_Forcing} with the carrier phase TKE held constant. The white surfaces are isocontours of $\phi=0.5$, representing interfaces, and the blue slices show the magnitude of velocity.}
    \label{fig:forcing-effect}
\end{figure}


The visualization of the two-phase turbulent system in the stationary state with different forcings is shown in Figure \ref{fig:forcing-effect}. Qualitatively, there is little difference in the dispersed phase structures in the stationary state with different forcings. This is consistent with the similar values of the mean surface energy achieved in the stationary state with different forcings, as seen in Figure \ref{fig:DiffForcings}(c).

\section{Analysis of the stationary states of two-phase isotropic turbulence}\label{sec:results}

In this section, we discuss the results of the forced simulations of two-phase homogeneous isotropic turbulence as a function of Weber number, void fraction, density ratio, and viscosity ratio, and present the observations and findings. 

\subsection{Effect of Weber number \label{sec:we_effect}}

By maintaining a unity density ratio, viscosity ratio, and a void fraction of $\langle \phi_3 \rangle=\langle \phi_d \rangle=0.0655$, we first study the effect of varying the Weber number by changing the surface tension values. The simulations that are compared here are listed as cases A, B1, and B2, in Table \ref{tab:TableCases}. Here in these simulations, both phases are forced to hold a constant value of the mean mixture TKE.  

\begin{figure}
    \centering 
    \includegraphics[width=\textwidth]{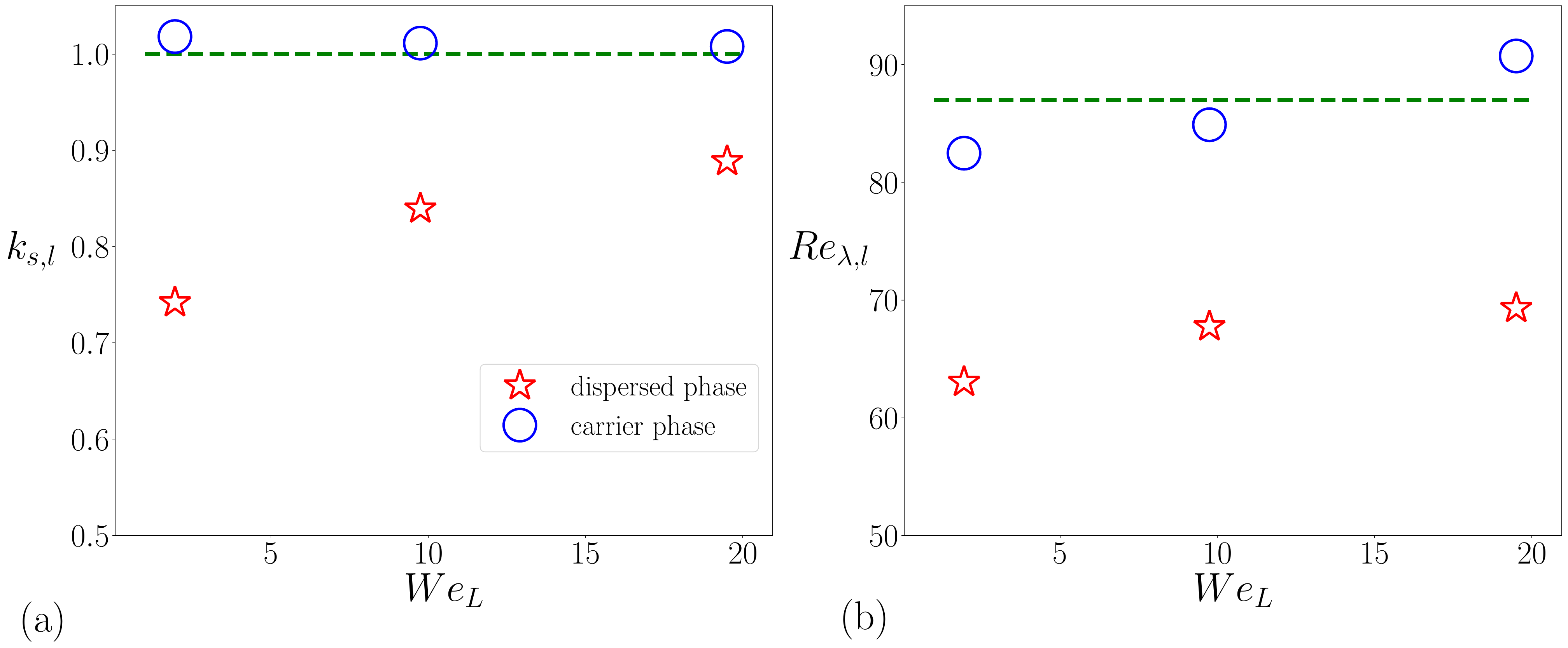}
    \caption{(a) The specific TKE of each phase, and (b) Taylor-scale Reynolds number, as a function of Weber number. The dashed line represents the reference single-phase value.}
    \label{fig:WeberKE_RE}
\end{figure}

Figure \ref{fig:WeberKE_RE}(a) shows the variation of specific TKE of each phase, defined as $k_{s,l}=\langle \phi_l u_{i,l} u_{i,l}/2 \rangle/\langle \phi_l \rangle$ for phase $l$, as the integral-scale Weber number $We_L$ is varied. Here, phase $2$ is the dispersed phase, and phase $1$ is the carrier phase. Since $\rho_2/\rho_1=1$ and $\mu_2/\mu_1=1$, and both phases are forced, we would expect $k_{s,l}$ to be $1$ (same) for both phases, but we see that the specific kinetic energy in the dispersed phase is less than in the carrier phase. This is probably because the interface limits the formation of large-scale eddies. Consequently, the Taylor-scale Reynolds number for the dispersed phase is lower than the carrier phase [Figure \ref{fig:WeberKE_RE}(b)], where the Taylor-scale Reynolds number is defined as $Re_{\lambda,l}={\{20/(3 \mu_l \epsilon_l)}\}^{0.5} k_l$ for phase $l$.
So, we postulate Hypothesis \ref{hyp:soft_wall} to explain these differences in the specific kinetic energies and Reynolds numbers of the dispersed and carrier phases.

\begin{figure}
    \centering 
    \includegraphics[width=0.8\textwidth]{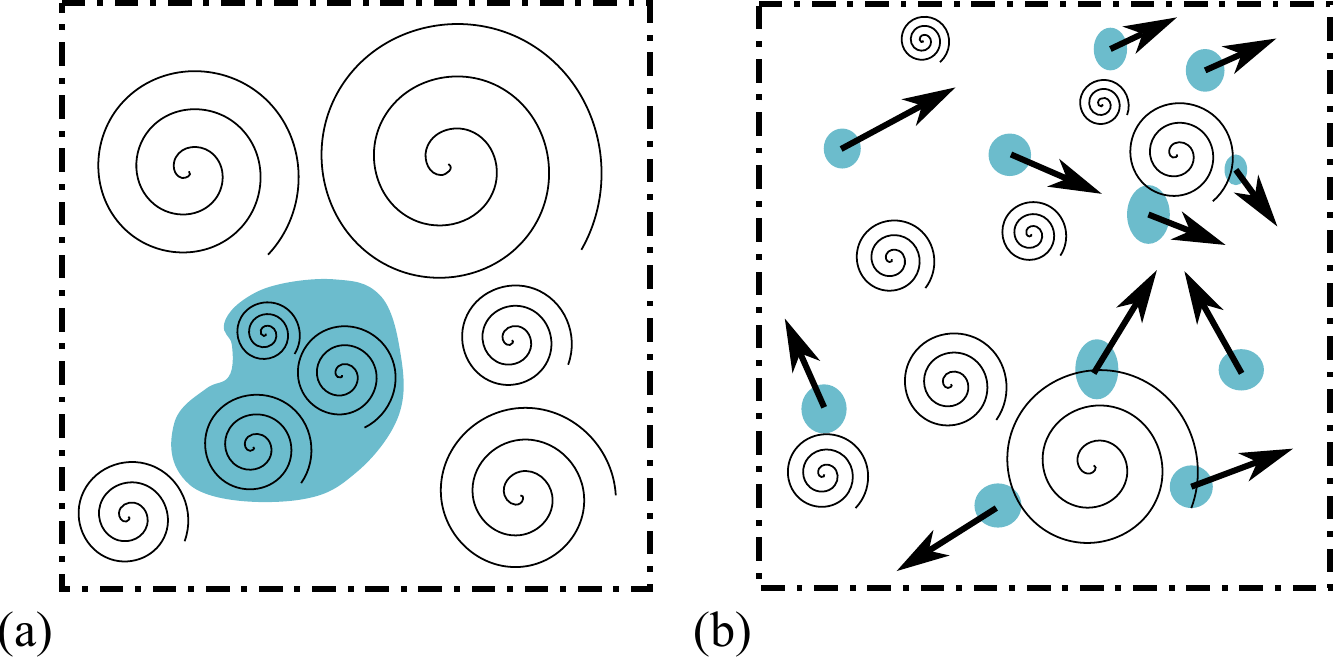}
    \caption{Schematic showing (a) the confinement effect of the capillary interface, and (b) the translational mode of energy stored in dispersed phase drops/bubbles in turbulence.  
    Here, the shaded region represents a drop/bubble.}
    \label{fig:Schematic}
\end{figure}

\newtheorem{hypothesis}{Hypothesis}[section]
\begin{hypothesis}
\textit{The capillary interface acts as a ``soft wall" and results in the confinement effect on the dispersed phase, inhibiting the formation of large-scale eddies (turbulent fluctuations larger than the drop/bubble size) as illustrated in Figure \ref{fig:Schematic}(a)}. 
\label{hyp:soft_wall}
\end{hypothesis}

Furthermore, it can be observed in Figure \ref{fig:WeberKE_RE}(a) that the specific TKE of the dispersed phase increases with $We_L$. This is a result of two possible effects. Firstly, as $We_L$ increases, the interface becomes more deformable and larger eddies can be accommodated within drops/bubbles (Hypothesis \ref{hyp:soft_wall}). 

Secondly, at higher values of $We_L$, there will be increased breakup resulting in higher number of smaller sub-Hinze drop/bubble formation (Section \ref{sec:size_dist}). Since the surface tension restoration force is stronger at sub-Hinze scales compared to the turbulent fluctuations, these bubbles/drops primarily store energy in translational modes than in turbulent fluctuations inside them [Hypothesis \ref{hyp:translation}]. Hence, an increase in the number of smaller drops/bubbles increases the amount of energy that is stored in the translational mode  of the dispersed phase cloud, as illustrated in Figure \ref{fig:Schematic}(b), contributing toward the overall specific kinetic energy of the dispersed phase. Consequently, the Taylor-scale Reynolds number of the dispersed phase also increases with $We_L$ [Figure \ref{fig:WeberKE_RE}(b)] because of these two effects. Note that the Taylor-scale Reynolds number here accounts for both modes of energy (turbulent fluctuations and translation) and its interpretation shifts with the change in phenomenology.
\begin{hypothesis}
\textit{The sub-Hinze bubbles/drops primarily store energy in translational modes.} 
\label{hyp:translation}
\end{hypothesis} 

One of these two factors could be dominant, or both factors could play a role, depending on the state of the system, and be responsible for the increase in the specific kinetic energy of the dispersed phase with an increase in the Weber number. There is a small increase in the Taylor-scale Reynolds number of the carrier phase with $We_L$, which is due to the bubble/drop-induced agitation in the carrier phase.

\subsubsection{Energy spectra}

\begin{figure}
    \centering
    \includegraphics[width=\linewidth]{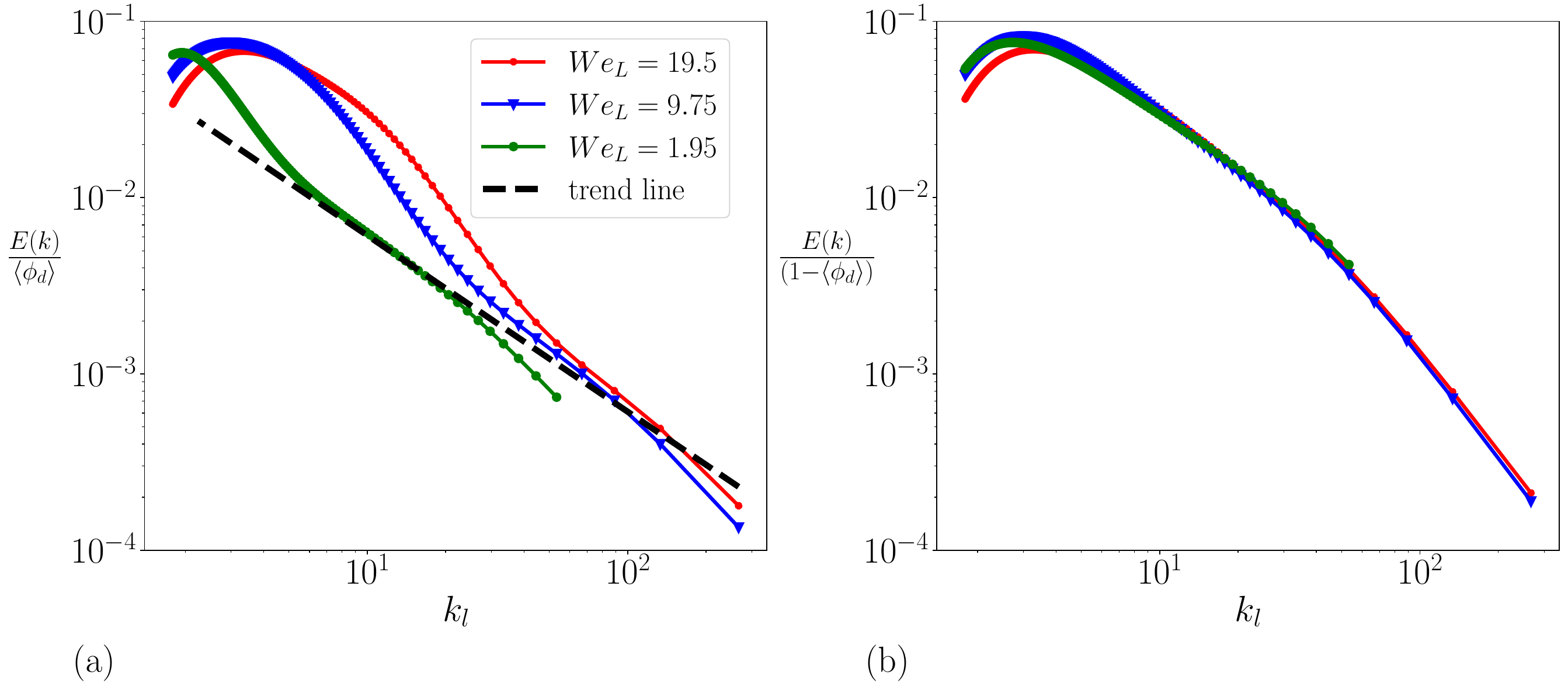}
    \caption{The energy spectrum normalized by the average void fraction in the (a) dispersed and (b) carrier phases of the two-phase turbulent flow for various Weber numbers, computed using the spatial filtering approach. The legend applies to both plots.}
    \label{fig:spectra-we}
\end{figure} 

Figure \ref{fig:spectra-we} shows the spectra, computed using the spatial filtering approach described in Section \ref{sec:spectrum}, in both the carrier and dispersed phases for all Weber numbers. The energy content in the dispersed phase is lower than in the carrier phase at all scales, which corroborates with the Hypothesis \ref{hyp:soft_wall} that the interface acts as a soft wall. Moreover, with increase in $We_L$, the energy content in the dispersed phase increases at all scales, consistent with the TKE in Figure \ref{fig:WeberKE_RE}, as the interface becomes more deformable and can accommodate more eddies in the dispersed phase. The carrier phase spectrum is mostly the same for all values of $We_L$, which is an expected behavior. 

In the dispersed phase, we observe two signatures\textemdash a high wavenumber trend (trend line) and a low wavenumber peak.
This behavior could be explained by the two internal modes of energy within the dispersed phase, i.e., turbulence within the dispersed phase at the small scales and the translational energy of the dispersed cloud (Hypothesis \ref{hyp:translation}) at the large scales. 
The separation between these two signatures in the plot also moves with $We_L$ indicating a shift in the Hinze scale and drop sizes. The low wavenumber peak is missing in the case of $We_L=1.95$ because of lack of breakup, denoting the absence of the translational mode of energy. 
It appears in Figure \ref{fig:spectra-we}(a) that this translational mode of energy is higher compared to the energy stored in turbulent fluctuations. This is because turbulence is more efficient in dissipating energy compared to the drag acting on the dispersed phase cloud to dissipate the translational mode of energy. This further explains the increase in specific TKE of the dispersed phase with $We_L$ in Figure \ref{fig:WeberKE_RE}(a).


\begin{figure}
    \centering
    \includegraphics[width=0.5\linewidth]{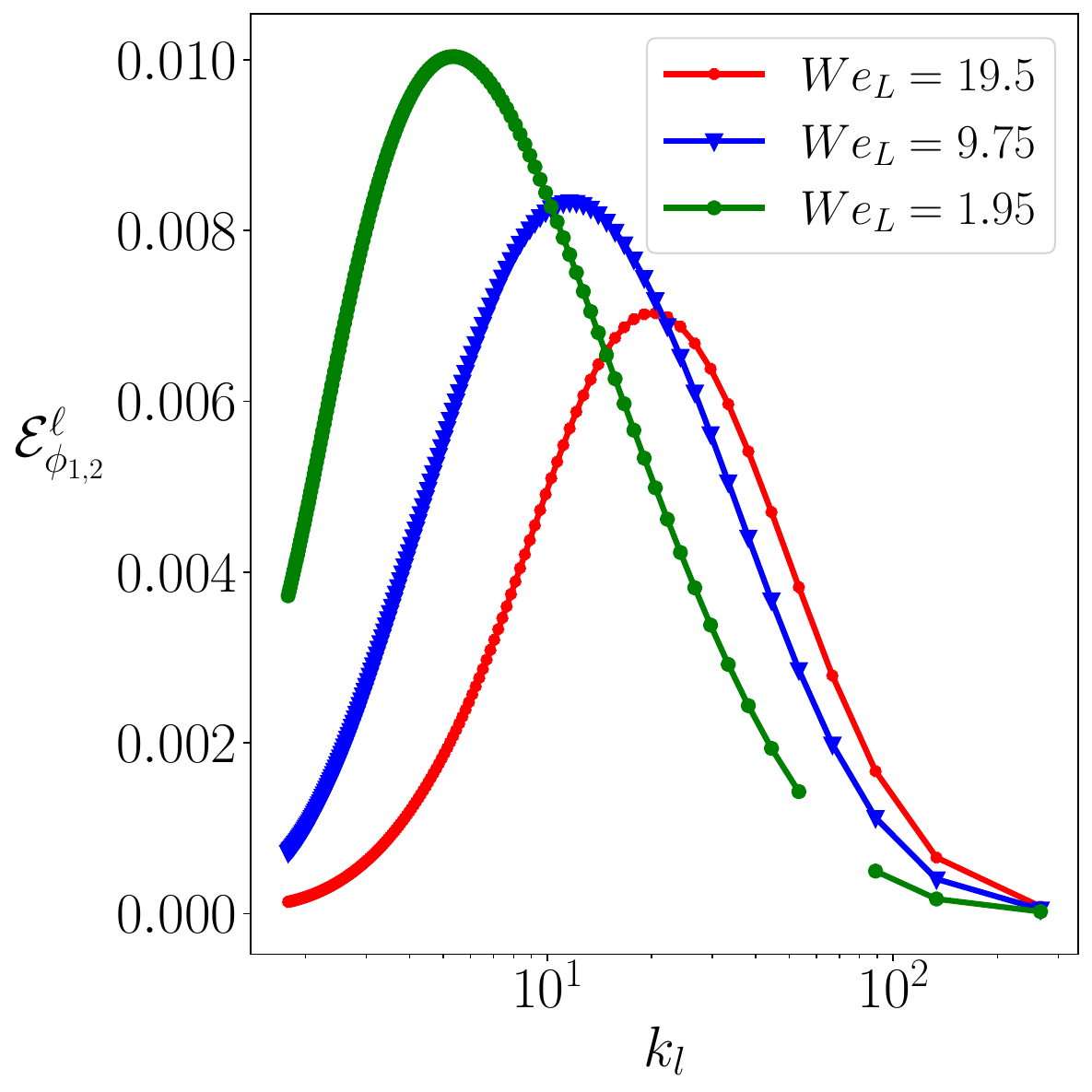}
    \caption{The cumulative mixture energy for various Weber numbers.}
    \label{fig:cum_energy-we}
\end{figure}
Figure \ref{fig:cum_energy-we} shows the cumulative mixture energy [relative velocity term in Eq. \eqref{eq:cum-energy}] as a function of wavenumber for various $We_L$. The peak here denotes the transition point from the dominant translation mode of energy (left of the peak) to the turbulent fluctuations (right of the peak). Consistent with the spectrum in Figure \ref{fig:spectra-we}(a), this peak shifts to the right with an increase in $We_L$.

\subsubsection{Size distribution \label{sec:size_dist}}
\begin{figure}
    \centering
    \includegraphics[width=0.85\textwidth]{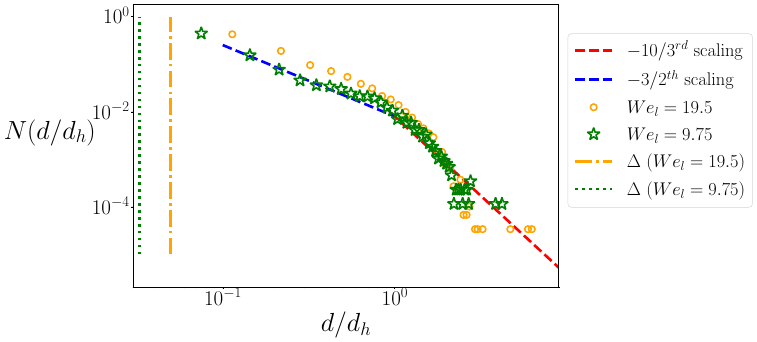}
    \caption{Effect of Weber number on the size distribution of the dispersed phase. The vertical lines represent grid sizes. The sizes are normalized by the Hinze scale, $d_h$.}
    \label{fig:WeberDSD}
\end{figure}

We plot the size distribution of the dispersed phase in the stationary state in Figure \ref{fig:WeberDSD} for $We_L=19.5$ (case A) and $9.75$ (case B1). The case B2 at $We_L=1.95$ is excluded here due to lack of breakup.  The sizes are normalized by the Hinze scale, defined as 
\begin{equation}
    \frac{d_h}{\eta} = 0.159 We_{crit}^{3/5} \frac{Re_{\lambda}^{3/2}}{We_L^{3/5}},
\end{equation}
where $We_{crit}=3$ is the critical Weber number, chosen based on \citet{riviere2021sub}. For both Weber numbers, the distribution follows a $-3/2^{th}$ power law for sub-Hinze scales and a $-10/3^{rd}$ power law for the super-Hinze scales, as seen before \citep{Deane2002, mukherjee2019droplet, riviere2021sub, crialesi2022modulation, roccon2023phase}; although a more steeper power law for super-Hinze scales has also been observed in the past \citep{li2017size} in more dynamic scenarios.
In the stationary size distribution of the dispersed cloud, we can see that for $We_L=19.5$, the number of sub-Hinze scale bubbles is higher compared to the case with $We_L=9.75$ due to the increase in breakup at higher Weber numbers. 
This increase in the number of sub-Hinze scale bubbles will contribute to the higher specific kinetic energy of the dispersed phase in Figure \ref{fig:WeberKE_RE}(a) in terms of translational mode (Hypothesis \ref{hyp:translation}), as described in Section \ref{sec:we_effect}. 

\begin{figure}
    \centering
    \includegraphics[width=\linewidth]{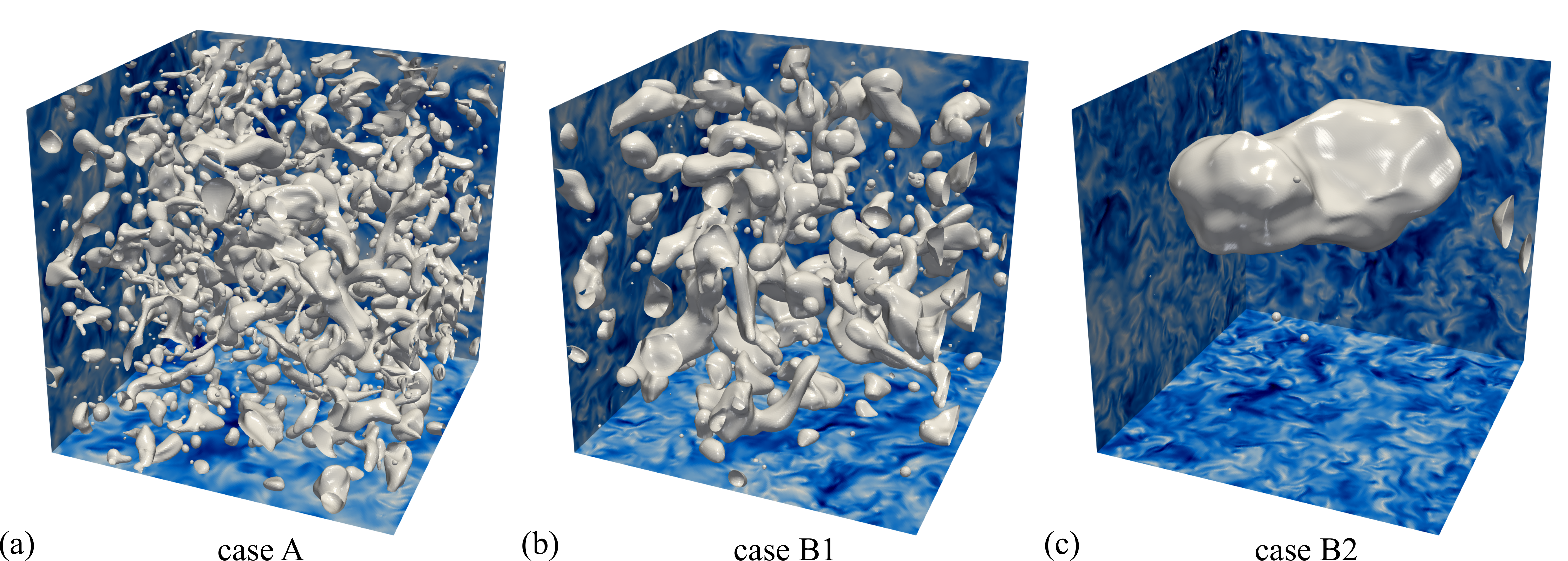}
    \caption{Visualization of the flow fields in stationary state for (a) case A at $We_L=19.5$, (b) case B1 at $We_L=9.75$, and (c) case B2 at $We_L=1.95$. The white surfaces are isocontours of $\phi=0.5$, representing interfaces, and the blue slices show the magnitude of velocity.}
    \label{fig:we-effect}
\end{figure}

The visualization of the two-phase turbulent system in the stationary state at different Weber numbers is shown in Figure \ref{fig:we-effect}. 
Consistent with the size distribution, there is a higher number of sub-Hinze scale drops/bubbles at $We_L=19.5$ than at $We_L=9.75$ due to increased breakup. There is no breakup at $We_L=1.95$ since this falls into the sub-critical Weber number regime.

\subsection{Effect of void fraction \label{sec:void}}

To study the effect of void fraction, while maintaining unity density and viscosity ratios, we first quantify how the energy is distributed between the two phases as void fraction increases from $0.0655$ to approximately $0.5$ (cases A, C1-C3), reaching a symmetric binary system. In this section, both the phases are forced using Eq. \eqref{Const_TKE_Forcing} to maintain a constant mean mixture TKE. 

\begin{figure}
    \centering
    \includegraphics[width=\textwidth]{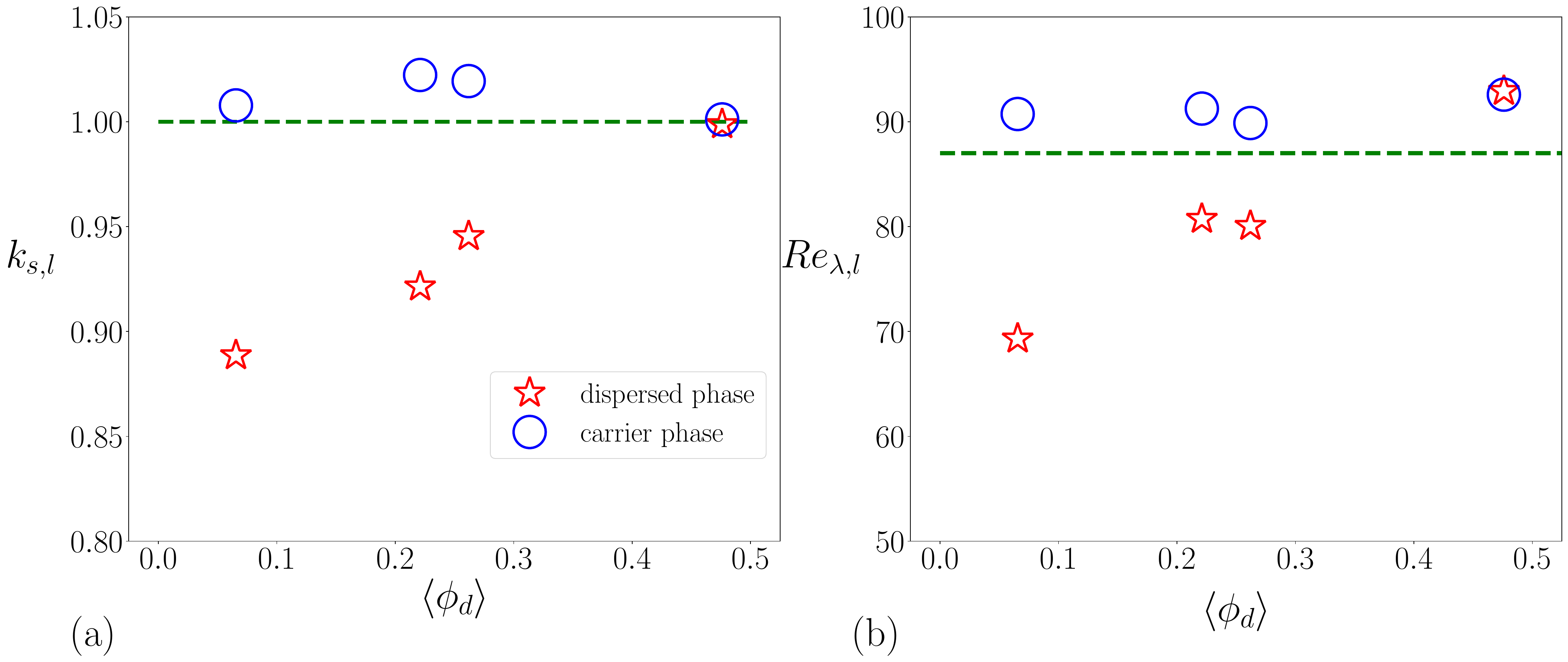}
    \caption{Effect of the void fraction on (a) the specific TKE of the dispersed and carrier phases, and (b) Taylor-scale Reynolds number of the dispersed and carrier phases, when both phases are forced simultaneously to maintain a constant value of the mixture TKE. The dashed line represents the reference single-phase value.}
    \label{fig:VoidKE_RE}
\end{figure}

Figure \ref{fig:VoidKE_RE}(a) shows the variation of the specific TKE of each phase, $\overline{k_s}$, as a function of the dispersed phase void fraction. 
We find that the specific kinetic energy of the dispersed phase increases approximately linearly with the increase in void fraction, reaching that of the ``carrier" phase as $\langle\phi_d\rangle\rightarrow0.5$, since there is no longer a distinction between the two phases of the emulsion at $\langle\phi_d\rangle = 0.5$. This is probably because of the increase in the size of bubbles/drops as the void fraction of the dispersed phase increases [this increase in the size of bubbles/drops as void fraction increases was also seen by \citet{crialesi2022modulation}]. Larger bubbles/drops can now accommodate larger eddies, which is consistent with the Hypothesis \ref{hyp:soft_wall}. 
\begin{figure}
    \centering
    \includegraphics[width=\linewidth]{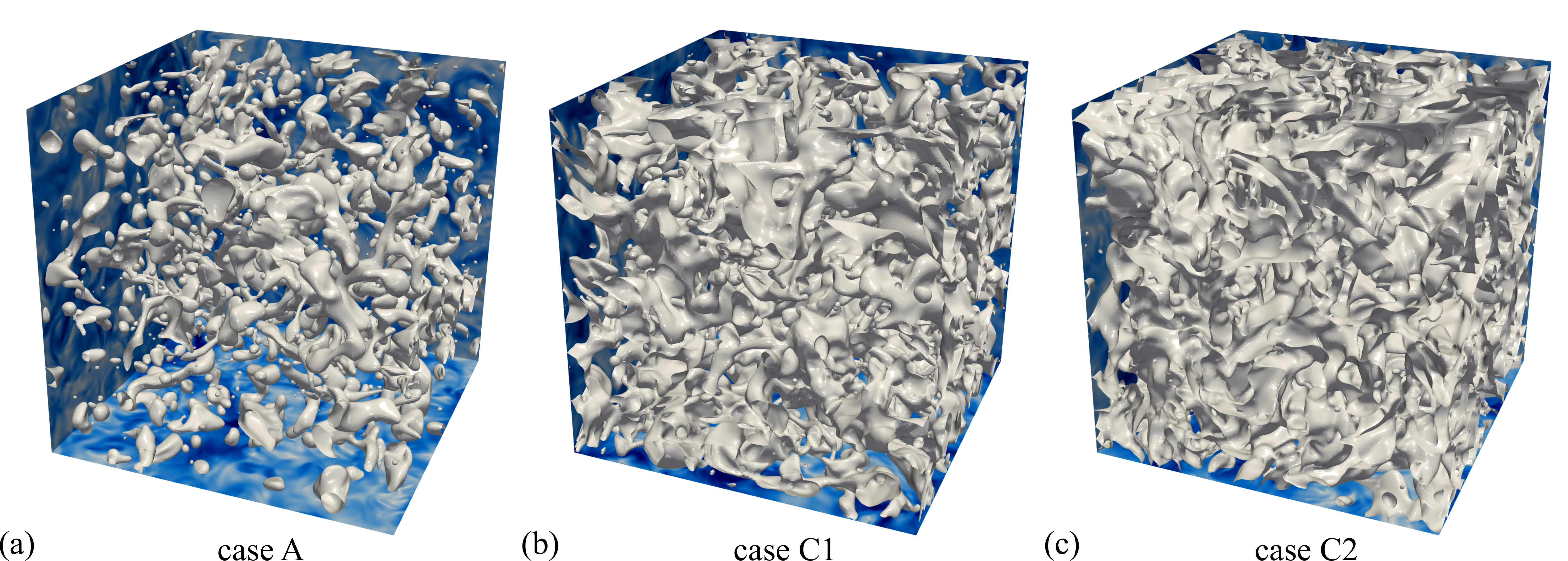}
    \caption{Visualization of the flow fields in stationary state for (a) case A at $\langle\phi_d\rangle=0.0655$, (b) case C1 at $\langle\phi_d\rangle=0.221$, and (c) case C2 at $\langle\phi_d\rangle=0.476$. The white surfaces are isocontours of $\phi=0.5$, representing interfaces, and the blue slices show the magnitude of velocity.}
    \label{fig:void-effect}
\end{figure}

The visualization of the two-phase turbulent mixture in a stationary state for various values of the void fraction is shown in Figure \ref{fig:void-effect}. As void fraction increases and reaches a value of $\langle \phi_d \rangle \rightarrow 0.5$, a symmetric binary mixture, the drops/bubbles grow in size and form a dense suspension.
By increasing the void fraction of phase 2 past $0.5$ (case C3 in Table \ref{tab:TableCases}), we find that the dispersed phase and carrier phase roles of the two phases (phase 2 and phase 1) switch and the specific kinetic energy of phase 1 decreases because it now becomes a dispersed phase (it was a carrier phase in other cases)
This confirms that the reduction in the specific kinetic energy observed here is not tied to any particular phase, but is indeed due to the dispersed nature of the phase. This reaffirms the inhibition of the turbulent fluctuations within the dispersed phase due to the interaction with the interface, which behaves as a ``soft wall" (Hypothesis \ref{hyp:soft_wall}). 

Similar to the specific TKE, the Taylor-scale Reynolds number, $\overline{Re}_{\lambda}$, of the dispersed phase increases approximately linearly with the increase in void fraction [Figure \ref{fig:VoidKE_RE}(b)], reaching that of the ``carrier" phase as $\langle \phi_d \rangle \rightarrow 0.5$. This is essentially a direct consequence of the increase of the specific TKE.


\subsubsection{Energy spectra}

\begin{figure}
    \centering
    \includegraphics[width=\linewidth]{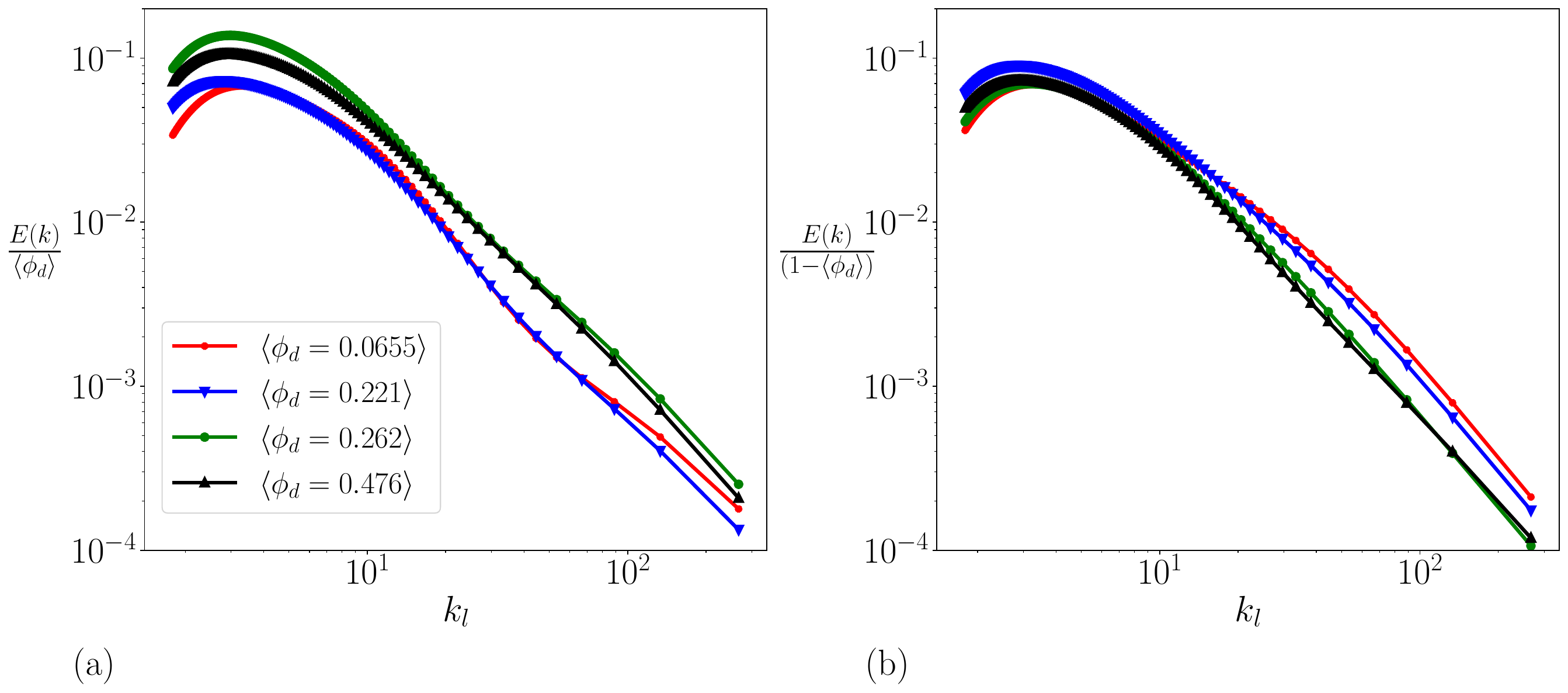}
    \caption{The energy spectrum normalized by the average void fraction in the (a) dispersed and (b) carrier phases of the two-phase turbulent flow for various void fractions, computed using the spatial filtering approach. The legend applies to both plots.}
    \label{fig:spectra-void}
\end{figure}

Figure \ref{fig:spectra-void} shows the spectra, computed using the spatial filtering approach described in Section \ref{sec:spectrum}, in both the carrier and dispersed phases for all values of void fractions. 
At lower values of $\langle \phi_d \rangle$, the energy content in the dispersed phase is lower than in the carrier phase at all scales, which corroborates with the Hypothesis \ref{hyp:soft_wall} that the interface acts as a soft wall. Moreover, with increase in $\langle \phi_d \rangle$, the energy content in the dispersed phase increases at all scales, consistent with the specific TKE in Figure \ref{fig:VoidKE_RE}, due to the increase in the size of the drops/bubbles and approaches that of the carrier phase for the higher $\langle \phi_d \rangle$ as the system approaches a symmetric binary system. The signature of the two energy modes (translation and turbulent fluctuations) seen for $\langle \phi_d \rangle = 0.0655$ also disappears as $\langle \phi_d \rangle \rightarrow 0.5$. The carrier phase spectrum is mostly the same for all values of $\langle \phi_d \rangle$, which is an expected behavior, except for lower values of $\langle \phi_d \rangle$ at small scales where there is a small increase in the energy probably due to the agitation effect of the dispersed phase cloud on the carrier phase.

\subsubsection{Interfacial area and surface energy}

\begin{figure}
    \centering
    \includegraphics[width=0.48\textwidth]{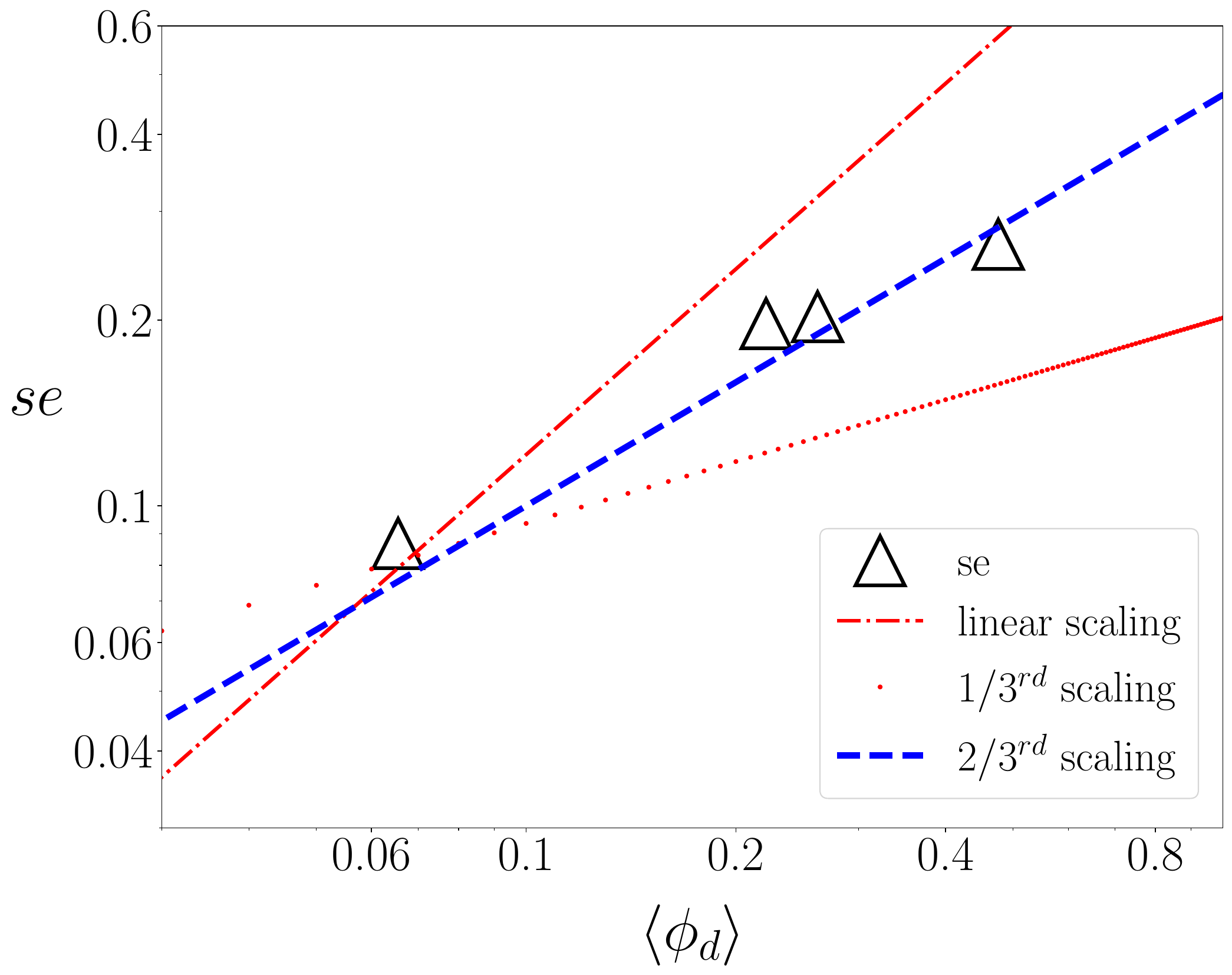}
    \caption{Effect of the void fraction on the mean surface energy of the two-phase mixture, when both the phases are forced to maintain a constant value of the mean mixture TKE.}
    \label{fig:VoidSE}
\end{figure}

Next, we compute the interfacial area and surface energy for cases A, C1-C3. Figure \ref{fig:VoidSE} shows the variation of the total surface energy of the two-phase mixture as a function of the dispersed phase void fraction. It appears that the mean surface energy and the mean interfacial area follows a power law
\begin{equation}
    se \sim \langle A \rangle \sim \langle\phi_d\rangle^{2/3},
\end{equation}
where $\langle A \rangle$ is the mean surface area of the dispersed phase. This power law can be easily explained because $\langle\phi_d\rangle \sim \ell^3$, where $\ell$ is an average length scale of the diameter of the dispersed phase, and $\langle A \rangle \sim \ell^2$. Therefore, $\langle A \rangle \sim \langle\phi_d\rangle^{2/3}$ and $se=\sigma \langle A \rangle$. It is expected that this area scaling will deviate from this ideal value once more super-Hinze scale bubbles are present at higher $Re$ due to intermittency and fractality.

\subsection{Effect of density ratio \label{sec:density}}

\begin{figure}
    \centering
    \includegraphics[width=\textwidth]{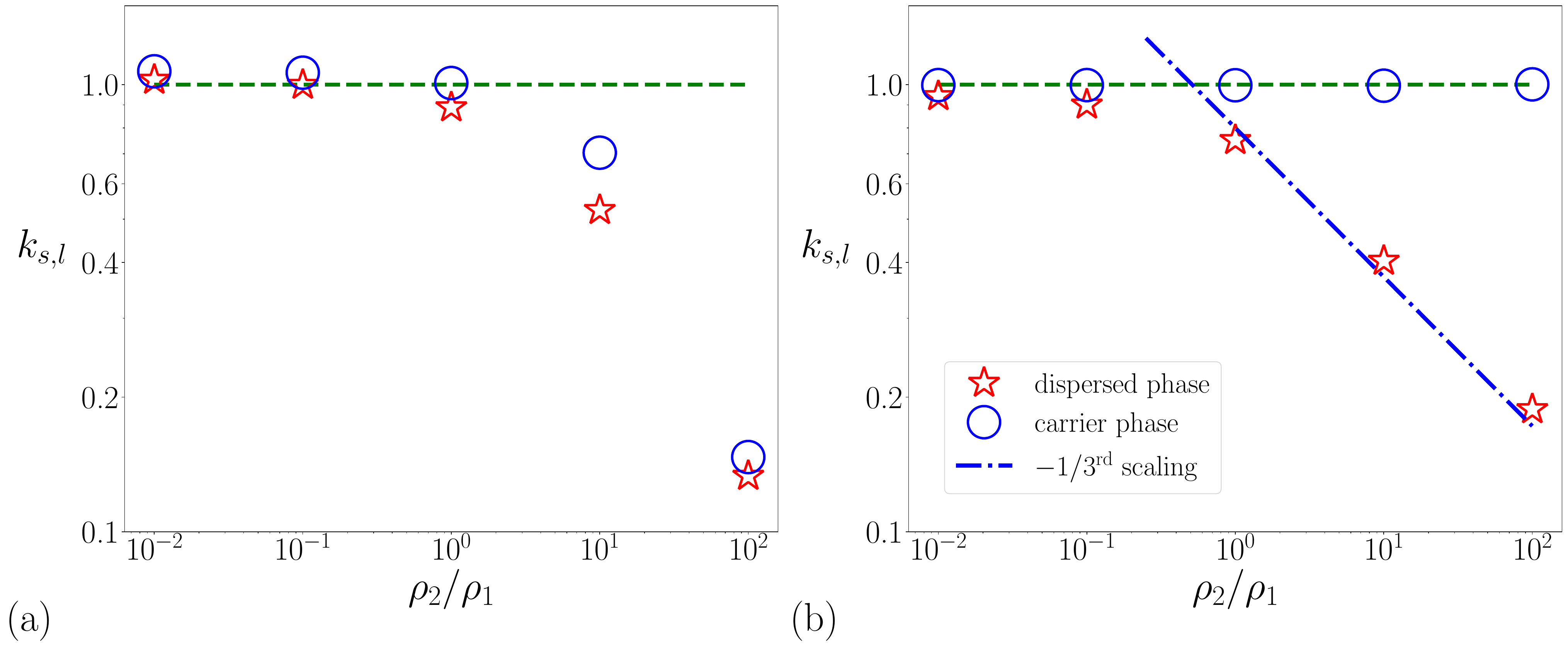}
    \caption{Effect of density ratio on the specific TKE of the phases for (a) constant mixture-TKE forcing, (b) constant carrier phase TKE forcing. The horizontal dashed line represents the reference single-phase value.} 
    
    \label{fig:DensityKE}
\end{figure}

We now examine the final steady state reached by a two-phase turbulent system as the dispersed phase changes from low-density bubbles to emulsions to high-density droplets. These correspond to cases A, D1-D4. Figure \ref{fig:DensityKE}(a) shows that if we force both phases simultaneously, a large variation in the specific energy distribution of the two phases does not exist until we transition to droplets. For droplets, simultaneously forcing both phases dumps a lot of kinetic energy into the droplets by virtue of their increased density, while draining the carrier phase considerably.
This is considered not physically relevant nor useful for two reasons. One, the turbulent intensity, $u'\sim \sqrt{2/3\overline{k}_s}$, of the carrier phase changes [Figure \ref{fig:DensityKE}(a)] as the density of the droplets changes, changing the turbulent state of the system. Second, the large-scale energy injection mechanisms in realistic engineering/natural applications likely act only on the carrier phase. 
So it might be more meaningful to hold the kinetic energy of the carrier phase constant, by forcing only the carrier phase, as we vary the density ratio. 

Figure \ref{fig:DensityKE}(b) shows the specific kinetic energy variation while holding the carrier phase TKE constant. There is a stark difference between the behavior of bubbles and droplets. For bubbles, their specific kinetic energy is roughly constant across two orders of magnitude of the density ratio. However, as soon as the dispersed phase changes to droplets, its specific kinetic energy decays following a $\sim(\rho_2/\rho_1)^{-1/3}$ 
power-law. 



\begin{figure}
    \centering
    \includegraphics[width=\linewidth]{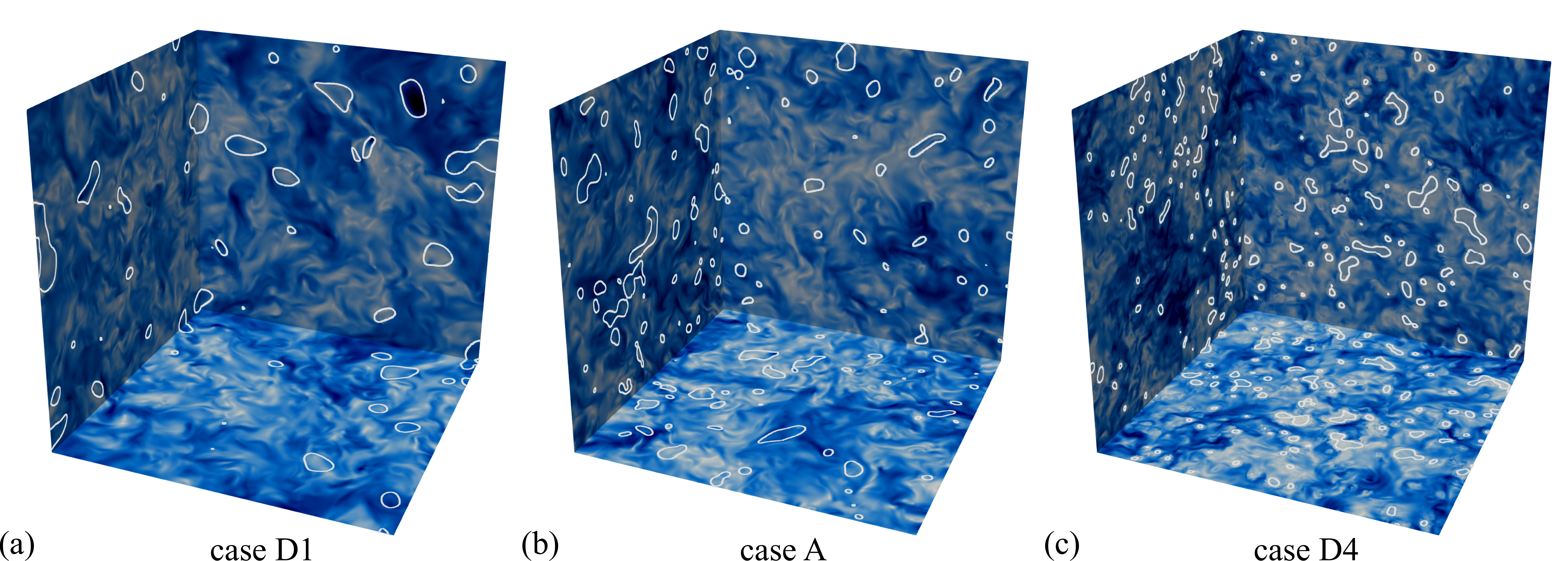}
    \caption{Visualization of the velocity fields in stationary state for (a) case D1 at $\rho_2/\rho_1 = 0.01$, $\mu_2/\mu_1 = 1$, (b) case A at $\rho_2/\rho_1 = 1$, $\mu_2/\mu_1 = 1$, and (c) case D4 at $\rho_2/\rho_1 = 100$, $\mu_2/\mu_1 = 1$. The white lines are isocontours of $\phi=0.5$, representing interfaces.}
    \label{fig:vel_density-effect}
\end{figure}

Balancing the tangential forces at the interface, one can relate the shear acting inside the dispersed phase with that in the carrier phase. Since the viscosity ratio is 1 for all the cases here, the shear from the carrier phase is transferred perfectly to drive the flow inside the dispersed phase. However, the difference in the inertia of the bubbles and droplets modifies the surrounding carrier phase through the normal stresses induced by collisions with turbulent eddies, leaving them unmodified in the case of bubbles and suppressing them in the case of droplets. Hence, the turbulence in the immediate vicinty of the bubbles remains strong, while it is heavily suppressed in the case of droplets. However, since the carrier phase remains continuously forced, this leads to stronger turbulence intensities in the carrier phase in the interstitial regions surrounding the droplets as seen in Figure \ref{fig:vel_density-effect}(c). This effect also explains the decay of the specific kinetic energy of the carrier phase when both phases are forced in Figure \ref{fig:DensityKE}(a). 

For the bubbles in cases D1 and D2, the kinematic viscosity, $\nu = \mu/\rho$, in the dispersed phase is higher than in the carrier phase due to lower density in the dispersed phase. This higher kinematic viscosity, which is essentially the momentum diffusivity, enhances the momentum exchange between the accelerations caused at the interface by the carrier phase and the interior bulk of the dispersed phase, bringing the dispersed phase in equilibrium with the carrier phase rapidly. This explains the constant specific kinetic energy for over two orders of magnitude of density ratio for bubbles, seen in Figure \ref{fig:DensityKE}(b). This observation is also consistent with the velocity field seen inside the dispersed phase for bubbles in Figure \ref{fig:vel_density-effect}(a); the velocity field inside the dispersed phase appears to be unaffected. Fluctuations inside droplets are weak due to their smaller sizes, as well as the depleted carrier phase fluctuations surrounding them.

\subsubsection{Energy spectra}

\begin{figure}
    \centering
    \includegraphics[width=\linewidth]{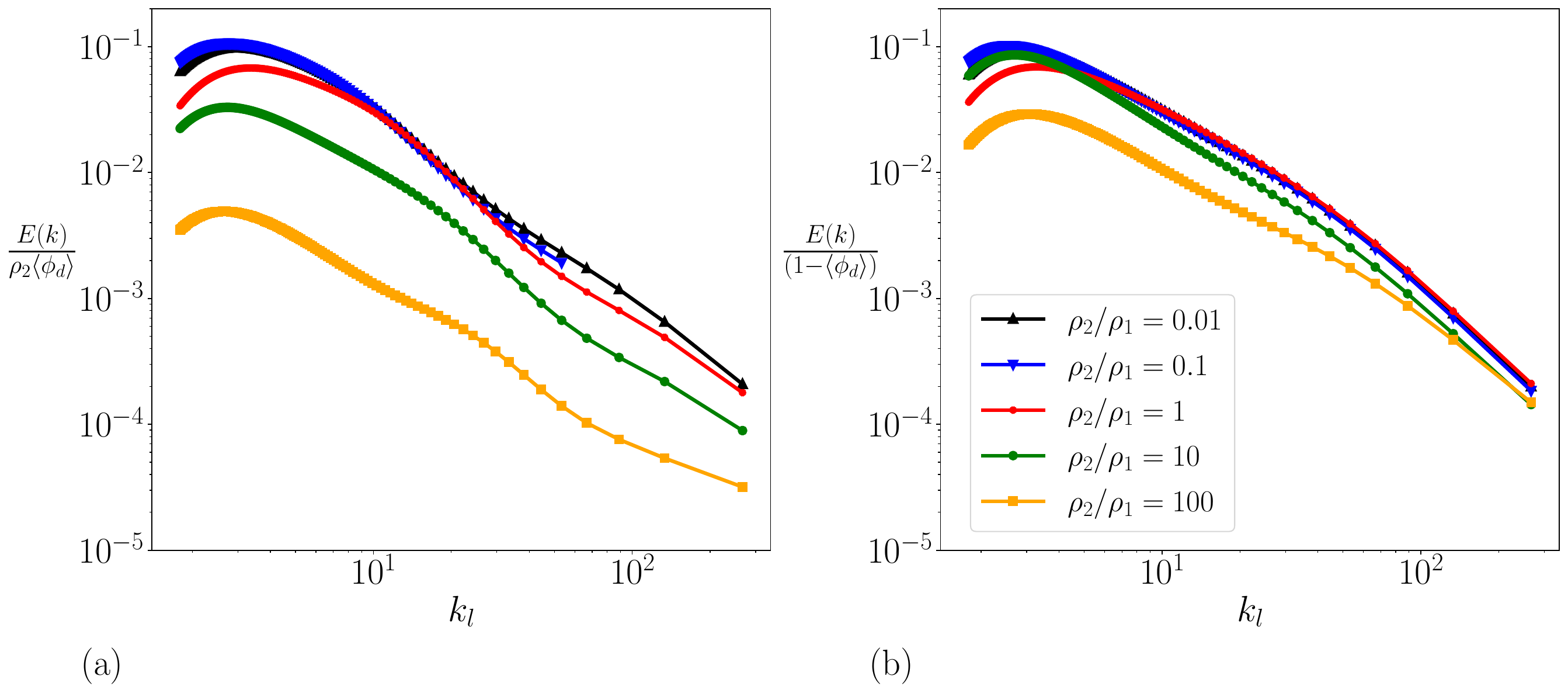}
    \caption{The energy spectrum normalized by the average void fraction in the (a) dispersed phase, compensated with the density of the dispersed phase, and (b) carrier phases of the two-phase turbulent flow for various density ratios, computed using the spatial filtering approach. The legend applies to both plots.}
    \label{fig:spectra-dr}
\end{figure}

Figure \ref{fig:spectra-dr} shows the spectra, computed using the spatial filtering approach described in Section \ref{sec:spectrum}, in both the carrier and dispersed phases for all the density ratios. 
The energy content in the dispersed phase is lower than in the carrier phase at all scales, which corroborates with the Hypothesis \ref{hyp:soft_wall} that the interface acts as a soft wall. 
Moreover, the spectrum in the dispersed phase is similar for bubbles (density ratio lower than 1) and emulsions (unity density), but the specific energy content for drops (density ratio higher than 1) is lower at all scales. This is consistent with the specific TKE in Figure \ref{fig:DensityKE}(b), and is due to the effect of momentum diffusivity and the inertia, as described in Section \ref{sec:density}. The carrier phase spectrum is mostly the same for all values of density ratio, which is an expected behavior, except for the case of high-density ratio droplets where there is some drainage of energy at large scales due to the inertia effect of droplets on the carrier phase as also seen in Figure \ref{fig:vel_density-effect}(c).

\subsection{Effect of viscosity ratio \label{sec:viscosity}}

\begin{figure}
    \centering
    \includegraphics[width=0.5\textwidth]{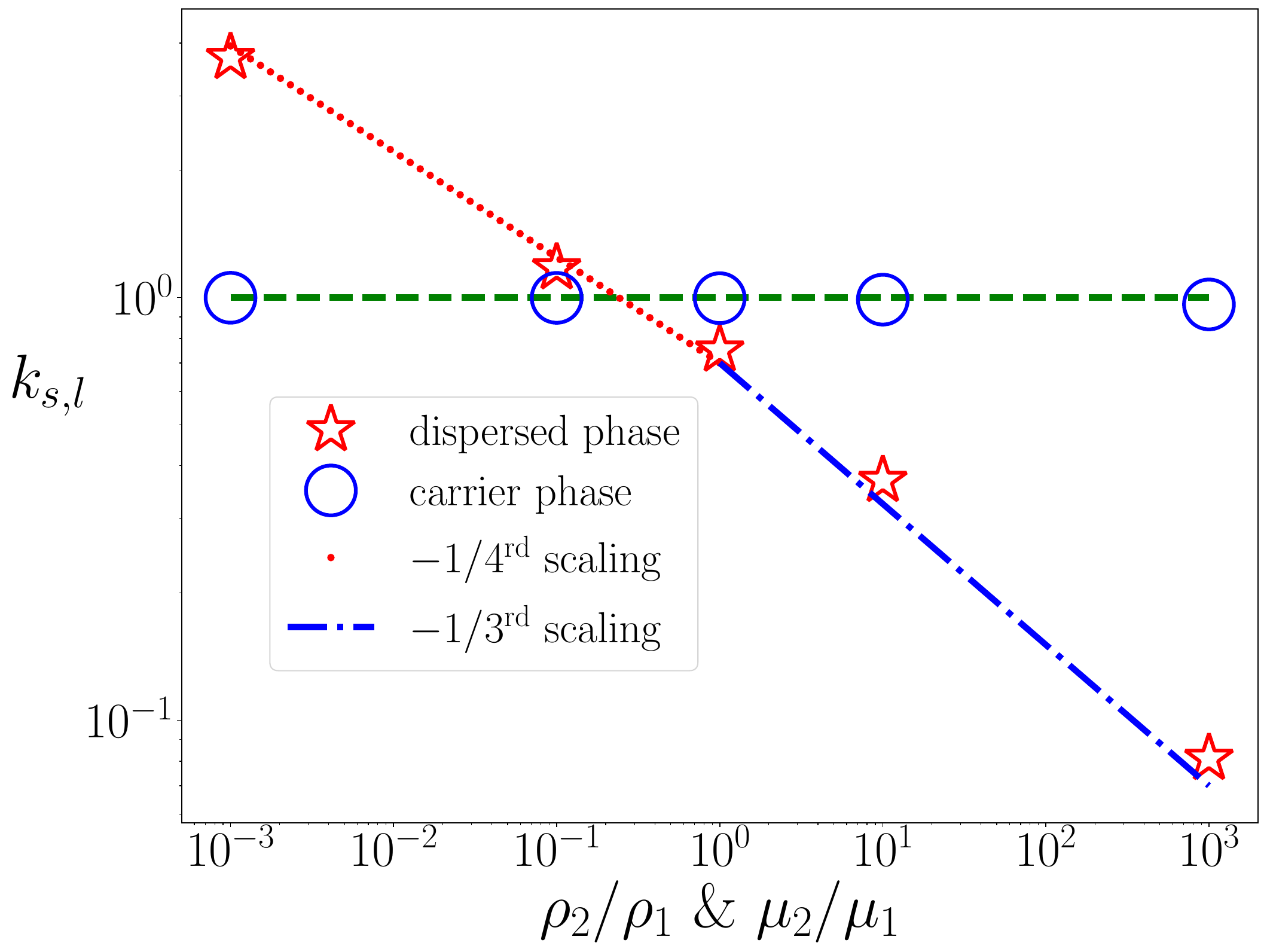}
    \caption{Effect of varying viscosity ratio, while holding kinematic viscosity constant, on the specific TKE of the phases. Here, the carrier phase is forced to hold a constant value of the carrier phase TKE. The horizontal dashed line represents the reference single-phase value.}
    \label{fig:ViscosityKE}
\end{figure} 

Finally, we examine the effect of varying the dynamic viscosity ratio between the dispersed and carrier phases. To isolate the effect of varying the viscosity ratio only, i.e., 
the velocity gradient jump on either side of the interface, while not influencing the momentum diffusivity inside each of the phases, we vary the density ratio simultaneously such that the kinematic viscosity remains constant in either phase. These correspond to cases A, E1-E4. Figure \ref{fig:ViscosityKE} shows the specific kinetic energy variation in the dispersed and carrier phases. Unlike Section \ref{sec:density}, the specific kinetic energy decays for both bubbles and drops as density and viscosity ratios increase, albeit with slightly different power laws. 

\begin{figure}
    \centering
    \includegraphics[width=\linewidth]{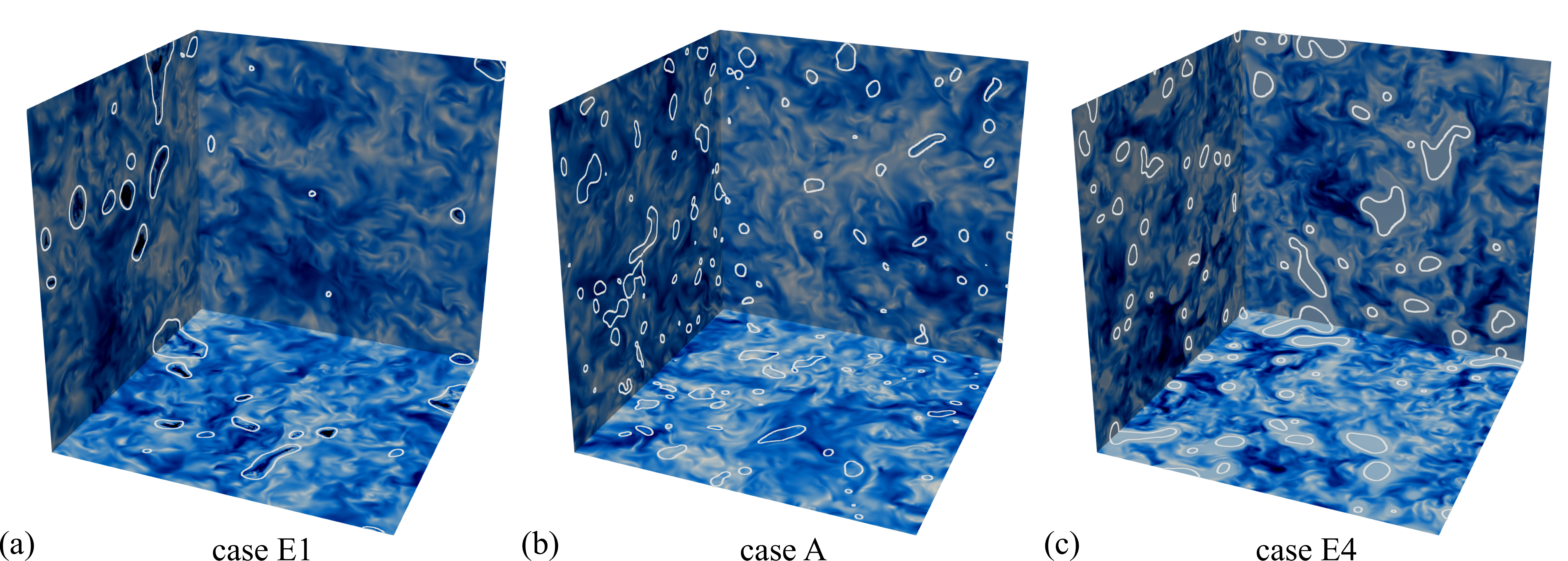}
    \caption{Visualization of the velocity fields in stationary state for (a) case E1 at $\rho_2/\rho_1 = 0.001$, $\mu_2/\mu_1 = 0.001$, (b) case A at $\rho_2/\rho_1 = 1$, $\mu_2/\mu_1 = 1$, and (c) case E4 at $\rho_2/\rho_1 = 1000$, $\mu_2/\mu_1 = 1000$. The white lines are isocontours of $\phi=0.5$, representing interfaces.}
    \label{fig:vel_viscosity-effect}
\end{figure}

For bubbles in cases E1 and E2, the specific kinetic energy in the dispersed phase is higher than in the carrier phase. Due to the tangential force balance at the interface between the two phases, the dynamic viscosity ratio causes a velocity gradient jump. Due to the low viscosity inside the bubbles, large velocity gradients are necessary to maintain the force balance. This, combined with the confinement effect of the interface, implies stronger internal fluctuations and hence a larger specific kinetic energy than the carrier phase as observed in Figure \ref{fig:vel_viscosity-effect}(a). Similar to cases D1 and D2, the lower inertia of the bubbles compared to the carrier phase lead to minor modulations of the turbulence on the carrier phase side. 

For the more viscous and denser drops, two effects occur simultaneously. The high viscosity ratio leads to weak velocity gradients inside the drops and hence weaker internal fluctuations. In conjunction, the higher inertia associated with the denser drops causes the suppression of the surrounding turbulence on the carrier phase side. This can be seen through the presence of boundary layers depleted of energy surrounding each of the drops in Figure \ref{fig:vel_viscosity-effect}(c). In contrast to the drops in cases D3 and D4, where only the second effect of the density ratio was active, the unity viscosity ratio in cases D3 and D4 led to stronger internal strain rates and hence more fragmentation, highlighted by the more numerous smaller droplets observed in Figure \ref{fig:vel_density-effect}(c) as opposed to the larger droplets observed in Figure \ref{fig:vel_viscosity-effect}(c). 


For naturally occurring air-water type two-phase systems, e.g., in oceanic applications, both viscosity and density ratios are non-unity values. Hence, the simulations presented in this section are most relevant to these realistic settings.  
Therefore, an air bubble suspended in water behaves fundamentally differently from a water droplet suspended in air. \textit{Bubbles have higher turbulent intensities inside them compared to the surrounding carrier phase, an effect primarily driven by the dynamic viscosity ratio. On the other hand, drops have lower turbulent intensities inside them compared to the surrounding carrier phase, which is a combined effect due to the viscosity ratio and in the increased inertia.}



\section{Summary and conclusions}\label{conclusions}

In this work, we performed forced numerical simulations of two-phase homogeneous isotropic turbulence and studied the statistical behavior of two-phase turbulent mixtures in a stationary state. We performed simulations for various density and viscosity ratios (representing bubbles, drops, and emulsions), Weber number, void fraction, type of forcing, and studied the effect of these on the global statistical properties and the phenomenological behavior of the two-phase mixture. 

We formulated three forcing schemes along with the use of a proportional controller that can hold at a constant value of (a) the turbulent kinetic energy (TKE) of the mixture (both phases forced), (b) the total energy, sum of turbulent kinetic energy and surface energy, of the mixture (both phases forced), and (c) the individual phase turbulent kinetic energy (typically carrier phase forced). When the contribution of the surface energy to the total energy of the system is negligible, there will be minimal differences between maintaining a constant value of TKE and the total energy. When the density ratio is unity, forcing both phases or only the carrier phase also leads to a similar stationary state. However, for non-unity density ratios, forcing only the carrier phase is more reasonable, particularly in the case of droplets. This is because simultaneously forcing both phases dumps higher kinetic energy into the droplets because of their increased density while draining the carrier phase, altering the turbulent intensity in the carrier phase and changing the turbulent state of the system. 

Forcing the two-phase turbulent system removes the arbitrariness associated with initializing the second phase in the simulation. Three different choices of the initial velocity field inside the dispersed phase, including (a) zero velocity, (b) retaining the same velocity as in the single-phase case, and (c) rescaling the velocity to maintain same kinetic energy when the density of the dispersed phase is different from the carrier phase, were tested with and without forcing. 
For the decaying case, these simulations showed that the time evolution of the two-phase turbulent system is heavily influenced by the choice of the initial condition, bringing arbitrariness into the system dynamics.  
For the forced simulations, all three choices of the initial condition reached the same stationary state, removing the arbitrariness in the system state. Furthermore, retaining the same velocity in the dispersed phase reached the stationary state the fastest and is, therefore, the recommended approach out of the three choices. Although the forced simulations reach the same stationary state, the system's evolution from the initial condition to the stationary state is still dependent on the choice of the initial condition and can last up to a few eddy turnover times. Therefore, simulations should be run long enough to flush out these initial transients.   

With these best practices on the choice of the forcing and the initial condition for two-phase turbulence simulations, forced numerical simulations of two-phase homogeneous isotropic turbulence were performed. It was found that the interface behaves as a ``soft wall" and inhibits the formation of larger eddies (turbulent fluctuations) within the dispersed phase. This hypothesis was developed by looking at the specific turbulent kinetic energy variation and Taylor-scale Reynolds number variation as a function of void fraction and Weber number. The specific turbulent kinetic energy and the Taylor-scale Reynolds number in the dispersed phase increased with an increase in void fraction and Weber number, indicating that larger eddies can be accommodated in larger bubbles/drops and when the interface is more deformable, respectively. Moreover, with an increase in Weber number, breakup increases, resulting in a higher number of sub-Hinze scale drops/bubbles. This potentially also increases the energy stored in the translational mode of the kinetic energy, which is the dominant mode of energy for sub-Hinze scale bubbles/drops.  

We also proposed using a spatial-filtering approach to compute spectra of the dispersed and carrier phases that is applicable for the general case of incompressible/compressible two-phase flows. A standard Fourier spectrum is not applicable when there are variations in the viscosity between the two phases due to the discontinuity of the velocity derivatives at the interface and when there are variations in the density between the two phases due to the cubic rather than quadratic nature of energy. While wavelet spectra can potentially remedy the viscosity issue by decomposing spatial regions into carrier, dispersed, and interaction phases, they are limited to quadratic definitions of energy like the Fourier spectrum, and can only report the dispersed phase spectrum for scales below the average size of the dispersed phase. The spatial filtering-approach can handle cubic energy and seamlessly transitions from scales below the average size of the dispersed phase, to scales between disjoint regions and beyond, where the interpretation changes from Lagrangian entities to an Eulerian field.
Using this approach, the spectra were computed for various void fractions, Weber numbers, and density ratios. 
The dispersed phase energy spectrum showed increase in energy at all scales as void fraction and Weber number increased, further corroborating with the hypothesis that interface behaves as a soft wall. The spectra also showed two peaks, which is the signature of the two energy modes (turbulent fluctuations and translational modes) in the in the dispersed phase. The dispersed phase was more efficient at storing energy in the translational energy mode. The carrier phase spectrum was invariant to the variation in the parameters as expected.

When void fraction of the two-phase mixture was varied, it was found that the global surface energy and the interfacial area of the two-phase mixture follow a $2/3^{rd}$ power-law of void fraction. Moreover, when the density ratio was varied, it was found that the specific kinetic energy follows a $-1/3^{rd}$ power-law of the density ratio for droplets (heavier dispersed phase). However, for bubbles, the variation in the specific kinetic energy with density ratio also depends on the viscosity ratio values. These differences are due to the varying momentum diffusivity that can play a significant role in bubbles due to low inertia. Finally, when both the density and viscosity ratios were varied such that momentum diffusivity ratio was unity, it was found that bubbles displayed stronger internal fluctuations compared to the carrier phase due to the enhanced shear and confinement effects. Droplets, on the other hand, behaved similarly to when only the density ratio was varied, albeit with less fragmentation, likely due to the weaker internal shear.

These interesting observations demonstrate that drops behave fundamentally differently from bubbles. Bubbles exhibit higher turbulent intensity compared to the surrounding fluid, whereas droplets exhibit lower turbulent intensity and behave ballistically.
Overall, this study provides insights into the complex dynamics of stationary two-phase turbulent flows and proposes best practices for their numerical simulations. This opens avenues for further studies into the development of subgrid-scale models of two-phase turbulent flows. 




\section*{Acknowledgements} 

This work was partially supported by the startup grant to S.~S.~J. from the George W. Woodruff School of Mechanical Engineering at Georgia Institute of Technology. S.~S.~J. also acknowledge partial funding support from Boeing Co. and GWW-GTRI Connect seed grant. A. E. acknowledges support from NASA Transformation Tools and Technologies Grant \#80NSSC20M0201.
This research used resources of the Oak Ridge Leadership Computing Facility, which is a DOE Office of Science User Facility supported under Contract DE-AC05-00OR22725. A preliminary version of this work has been published as a technical report in the annual publication of the Center for Turbulence Research \citep{jain_forced2023} and is available online\footnote{https://web.stanford.edu/group/ctr/ResBriefs/2023/12\_Jain.pdf}.

\bibliographystyle{model1-num-names}
\bibliography{sample}

\begin{thebibliography}{53}
\expandafter\ifx\csname natexlab\endcsname\relax\def\natexlab#1{#1}\fi
\providecommand{\bibinfo}[2]{#2}
\ifx\xfnm\relax \def\xfnm[#1]{\unskip,\space#1}\fi
\bibitem[{Liao and Lucas(2009)}]{liao2009literature}
\bibinfo{author}{Y.~Liao}, \bibinfo{author}{D.~Lucas},
\newblock \bibinfo{title}{A literature review of theoretical models for drop and bubble breakup in turbulent dispersions},
\newblock \bibinfo{journal}{Chemical Engineering Science} \bibinfo{volume}{64} (\bibinfo{year}{2009}) \bibinfo{pages}{3389--3406}.
\bibitem[{Liao and Lucas(2010)}]{liao2010literature}
\bibinfo{author}{Y.~Liao}, \bibinfo{author}{D.~Lucas},
\newblock \bibinfo{title}{A literature review on mechanisms and models for the coalescence process of fluid particles},
\newblock \bibinfo{journal}{Chemical Engineering Science} \bibinfo{volume}{65} (\bibinfo{year}{2010}) \bibinfo{pages}{2851--2864}.
\bibitem[{Jain(2017)}]{jain2017flow}
\bibinfo{author}{S.~Jain},
\newblock \bibinfo{title}{Flow-induced breakup of drops and bubbles},
\newblock \bibinfo{journal}{arXiv e-prints}  (\bibinfo{year}{2017}) \bibinfo{pages}{arXiv--1701}.
\bibitem[{Ni(2024)}]{ni2024deformation}
\bibinfo{author}{R.~Ni},
\newblock \bibinfo{title}{Deformation and breakup of bubbles and drops in turbulence},
\newblock \bibinfo{journal}{Annual Review of Fluid Mechanics} \bibinfo{volume}{56} (\bibinfo{year}{2024}) \bibinfo{pages}{319--347}.
\bibitem[{Grabowski and Wang(2013)}]{grabowski2013growth}
\bibinfo{author}{W.~W. Grabowski}, \bibinfo{author}{L.-P. Wang},
\newblock \bibinfo{title}{Growth of cloud droplets in a turbulent environment},
\newblock \bibinfo{journal}{Annual review of fluid mechanics} \bibinfo{volume}{45} (\bibinfo{year}{2013}) \bibinfo{pages}{293--324}.
\bibitem[{Deike(2022)}]{deike2022mass}
\bibinfo{author}{L.~Deike},
\newblock \bibinfo{title}{Mass transfer at the ocean--atmosphere interface: the role of wave breaking, droplets, and bubbles},
\newblock \bibinfo{journal}{Annual Review of Fluid Mechanics} \bibinfo{volume}{54} (\bibinfo{year}{2022}) \bibinfo{pages}{191--224}.
\bibitem[{Wang et~al.(2007)Wang, Li, Zhang, Dong, and Eastoe}]{wang2007oil}
\bibinfo{author}{L.~Wang}, \bibinfo{author}{X.~Li}, \bibinfo{author}{G.~Zhang}, \bibinfo{author}{J.~Dong}, \bibinfo{author}{J.~Eastoe},
\newblock \bibinfo{title}{Oil-in-water nanoemulsions for pesticide formulations},
\newblock \bibinfo{journal}{Journal of colloid and interface science} \bibinfo{volume}{314} (\bibinfo{year}{2007}) \bibinfo{pages}{230--235}.
\bibitem[{Kilpatrick(2012)}]{kilpatrick2012water}
\bibinfo{author}{P.~K. Kilpatrick},
\newblock \bibinfo{title}{Water-in-crude oil emulsion stabilization: review and unanswered questions},
\newblock \bibinfo{journal}{Energy \& Fuels} \bibinfo{volume}{26} (\bibinfo{year}{2012}) \bibinfo{pages}{4017--4026}.
\bibitem[{Rosti et~al.(2019)Rosti, Ge, Jain, Dodd, and Brandt}]{rosti2019droplets}
\bibinfo{author}{M.~E. Rosti}, \bibinfo{author}{Z.~Ge}, \bibinfo{author}{S.~S. Jain}, \bibinfo{author}{M.~S. Dodd}, \bibinfo{author}{L.~Brandt},
\newblock \bibinfo{title}{Droplets in homogeneous shear turbulence},
\newblock \bibinfo{journal}{Journal of Fluid Mechanics} \bibinfo{volume}{876} (\bibinfo{year}{2019}) \bibinfo{pages}{962--984}.
\bibitem[{Ravelet et~al.(2011)Ravelet, Colin, and Risso}]{ravelet2011dynamics}
\bibinfo{author}{F.~Ravelet}, \bibinfo{author}{C.~Colin}, \bibinfo{author}{F.~Risso},
\newblock \bibinfo{title}{On the dynamics and breakup of a bubble rising in a turbulent flow},
\newblock \bibinfo{journal}{Physics of Fluids} \bibinfo{volume}{23} (\bibinfo{year}{2011}).
\bibitem[{Hinze(1955)}]{hinze1955fundamentals}
\bibinfo{author}{J.~O. Hinze},
\newblock \bibinfo{title}{Fundamentals of the hydrodynamic mechanism of splitting in dispersion processes},
\newblock \bibinfo{journal}{AIChE journal} \bibinfo{volume}{1} (\bibinfo{year}{1955}) \bibinfo{pages}{289--295}.
\bibitem[{Eastwood et~al.(2004)Eastwood, Armi, and Lasheras}]{eastwood2004breakup}
\bibinfo{author}{C.~D. Eastwood}, \bibinfo{author}{L.~Armi}, \bibinfo{author}{J.~Lasheras},
\newblock \bibinfo{title}{The breakup of immiscible fluids in turbulent flows},
\newblock \bibinfo{journal}{Journal of Fluid Mechanics} \bibinfo{volume}{502} (\bibinfo{year}{2004}) \bibinfo{pages}{309--333}.
\bibitem[{Kim and Moin(2020)}]{kim2020subgrid}
\bibinfo{author}{D.~Kim}, \bibinfo{author}{P.~Moin},
\newblock \bibinfo{title}{Subgrid-scale capillary breakup model for liquid jet atomization},
\newblock \bibinfo{journal}{Combustion Science and Technology} \bibinfo{volume}{192} (\bibinfo{year}{2020}) \bibinfo{pages}{1334--1357}.
\bibitem[{Chan et~al.(2021{\natexlab{a}})Chan, Johnson, and Moin}]{chan2021turbulenta}
\bibinfo{author}{W.~H.~R. Chan}, \bibinfo{author}{P.~L. Johnson}, \bibinfo{author}{P.~Moin},
\newblock \bibinfo{title}{The turbulent bubble break-up cascade. part 1. theoretical developments},
\newblock \bibinfo{journal}{Journal of Fluid Mechanics} \bibinfo{volume}{912} (\bibinfo{year}{2021}{\natexlab{a}}) \bibinfo{pages}{A42}.
\bibitem[{Chan et~al.(2021{\natexlab{b}})Chan, Johnson, Moin, and Urzay}]{chan2021turbulentb}
\bibinfo{author}{W.~H.~R. Chan}, \bibinfo{author}{P.~L. Johnson}, \bibinfo{author}{P.~Moin}, \bibinfo{author}{J.~Urzay},
\newblock \bibinfo{title}{The turbulent bubble break-up cascade. part 2. numerical simulations of breaking waves},
\newblock \bibinfo{journal}{Journal of Fluid Mechanics} \bibinfo{volume}{912} (\bibinfo{year}{2021}{\natexlab{b}}) \bibinfo{pages}{A43}.
\bibitem[{Qi et~al.(2022)Qi, Tan, Corbitt, Urbanik, Salibindla, and Ni}]{qi2022fragmentation}
\bibinfo{author}{Y.~Qi}, \bibinfo{author}{S.~Tan}, \bibinfo{author}{N.~Corbitt}, \bibinfo{author}{C.~Urbanik}, \bibinfo{author}{A.~K. Salibindla}, \bibinfo{author}{R.~Ni},
\newblock \bibinfo{title}{Fragmentation in turbulence by small eddies},
\newblock \bibinfo{journal}{Nature communications} \bibinfo{volume}{13} (\bibinfo{year}{2022}) \bibinfo{pages}{469}.
\bibitem[{Riviere et~al.(2021)Riviere, Mostert, Perrard, and Deike}]{riviere2021sub}
\bibinfo{author}{A.~Riviere}, \bibinfo{author}{W.~Mostert}, \bibinfo{author}{S.~Perrard}, \bibinfo{author}{L.~Deike},
\newblock \bibinfo{title}{Sub-hinze scale bubble production in turbulent bubble break-up},
\newblock \bibinfo{journal}{Journal of Fluid Mechanics} \bibinfo{volume}{917} (\bibinfo{year}{2021}) \bibinfo{pages}{A40}.
\bibitem[{Rivi{\`e}re et~al.(2022)Rivi{\`e}re, Ruth, Mostert, Deike, and Perrard}]{riviere2022capillary}
\bibinfo{author}{A.~Rivi{\`e}re}, \bibinfo{author}{D.~J. Ruth}, \bibinfo{author}{W.~Mostert}, \bibinfo{author}{L.~Deike}, \bibinfo{author}{S.~Perrard},
\newblock \bibinfo{title}{Capillary driven fragmentation of large gas bubbles in turbulence},
\newblock \bibinfo{journal}{Physical Review Fluids} \bibinfo{volume}{7} (\bibinfo{year}{2022}) \bibinfo{pages}{083602}.
\bibitem[{Vela-Mart{\'\i}n and Avila(2021)}]{vela2021deformation}
\bibinfo{author}{A.~Vela-Mart{\'\i}n}, \bibinfo{author}{M.~Avila},
\newblock \bibinfo{title}{Deformation of drops by outer eddies in turbulence},
\newblock \bibinfo{journal}{Journal of Fluid Mechanics} \bibinfo{volume}{929} (\bibinfo{year}{2021}) \bibinfo{pages}{A38}.
\bibitem[{Perrard et~al.(2021)Perrard, Rivi{\`e}re, Mostert, and Deike}]{perrard2021bubble}
\bibinfo{author}{S.~Perrard}, \bibinfo{author}{A.~Rivi{\`e}re}, \bibinfo{author}{W.~Mostert}, \bibinfo{author}{L.~Deike},
\newblock \bibinfo{title}{Bubble deformation by a turbulent flow},
\newblock \bibinfo{journal}{Journal of Fluid Mechanics} \bibinfo{volume}{920} (\bibinfo{year}{2021}) \bibinfo{pages}{A15}.
\bibitem[{Vela-Mart{\'\i}n and Avila(2022)}]{vela2022memoryless}
\bibinfo{author}{A.~Vela-Mart{\'\i}n}, \bibinfo{author}{M.~Avila},
\newblock \bibinfo{title}{Memoryless drop breakup in turbulence},
\newblock \bibinfo{journal}{Science Advances} \bibinfo{volume}{8} (\bibinfo{year}{2022}) \bibinfo{pages}{eabp9561}.
\bibitem[{Farsoiya et~al.(2023)Farsoiya, Liu, Daiss, Fox, and Deike}]{farsoiya2023role}
\bibinfo{author}{P.~K. Farsoiya}, \bibinfo{author}{Z.~Liu}, \bibinfo{author}{A.~Daiss}, \bibinfo{author}{R.~O. Fox}, \bibinfo{author}{L.~Deike},
\newblock \bibinfo{title}{Role of viscosity in turbulent drop break-up},
\newblock \bibinfo{journal}{Journal of Fluid Mechanics} \bibinfo{volume}{972} (\bibinfo{year}{2023}) \bibinfo{pages}{A11}.
\bibitem[{Dodd and Ferrante(2016)}]{dodd2016interaction}
\bibinfo{author}{M.~S. Dodd}, \bibinfo{author}{A.~Ferrante},
\newblock \bibinfo{title}{On the interaction of taylor length scale size droplets and isotropic turbulence},
\newblock \bibinfo{journal}{Journal of Fluid Mechanics} \bibinfo{volume}{806} (\bibinfo{year}{2016}) \bibinfo{pages}{356--412}.
\bibitem[{Crialesi-Esposito et~al.(2022)Crialesi-Esposito, Rosti, Chibbaro, and Brandt}]{crialesi2022modulation}
\bibinfo{author}{M.~Crialesi-Esposito}, \bibinfo{author}{M.~E. Rosti}, \bibinfo{author}{S.~Chibbaro}, \bibinfo{author}{L.~Brandt},
\newblock \bibinfo{title}{Modulation of homogeneous and isotropic turbulence in emulsions},
\newblock \bibinfo{journal}{Journal of Fluid Mechanics} \bibinfo{volume}{940} (\bibinfo{year}{2022}) \bibinfo{pages}{A19}.
\bibitem[{Crialesi-Esposito et~al.(2023)Crialesi-Esposito, Chibbaro, and Brandt}]{crialesi2023interaction}
\bibinfo{author}{M.~Crialesi-Esposito}, \bibinfo{author}{S.~Chibbaro}, \bibinfo{author}{L.~Brandt},
\newblock \bibinfo{title}{The interaction of droplet dynamics and turbulence cascade},
\newblock \bibinfo{journal}{Communications Physics} \bibinfo{volume}{6} (\bibinfo{year}{2023}) \bibinfo{pages}{5}.
\bibitem[{Begemann et~al.(2022)Begemann, Trummler, Trautner, Hasslberger, and Klein}]{begemann2022effect}
\bibinfo{author}{A.~Begemann}, \bibinfo{author}{T.~Trummler}, \bibinfo{author}{E.~Trautner}, \bibinfo{author}{J.~Hasslberger}, \bibinfo{author}{M.~Klein},
\newblock \bibinfo{title}{Effect of turbulence intensity and surface tension on the emulsification process and its stationary state—a numerical study},
\newblock \bibinfo{journal}{The Canadian Journal of Chemical Engineering} \bibinfo{volume}{100} (\bibinfo{year}{2022}) \bibinfo{pages}{3548--3561}.
\bibitem[{Krzeczek et~al.(2023)Krzeczek, Trummler, Trautner, and Klein}]{krzeczek2023effect}
\bibinfo{author}{O.~Krzeczek}, \bibinfo{author}{T.~Trummler}, \bibinfo{author}{E.~Trautner}, \bibinfo{author}{M.~Klein},
\newblock \bibinfo{title}{Effect of the density ratio on emulsions and their segregation: A direct numerical simulation study},
\newblock \bibinfo{journal}{Energies} \bibinfo{volume}{16} (\bibinfo{year}{2023}) \bibinfo{pages}{3160}.
\bibitem[{Lundgren(2003)}]{lundgren2003linearly}
\bibinfo{author}{T.~S. Lundgren},
\newblock \bibinfo{title}{Linearly forced isotropic turbulence},
\newblock \bibinfo{journal}{Center for Turbulence Research Annual Research Briefs 2003}  (\bibinfo{year}{2003}).
\bibitem[{Rosales and Meneveau(2005)}]{rosales2005linear}
\bibinfo{author}{C.~Rosales}, \bibinfo{author}{C.~Meneveau},
\newblock \bibinfo{title}{Linear forcing in numerical simulations of isotropic turbulence: Physical space implementations and convergence properties},
\newblock \bibinfo{journal}{Physics of fluids} \bibinfo{volume}{17} (\bibinfo{year}{2005}).
\bibitem[{Boukharfane et~al.(2021)Boukharfane, Er-Raiy, Parsani, and Chakraborty}]{boukharfane2021structure}
\bibinfo{author}{R.~Boukharfane}, \bibinfo{author}{A.~Er-Raiy}, \bibinfo{author}{M.~Parsani}, \bibinfo{author}{N.~Chakraborty},
\newblock \bibinfo{title}{Structure and dynamics of small-scale turbulence in vaporizing two-phase flows},
\newblock \bibinfo{journal}{Scientific reports} \bibinfo{volume}{11} (\bibinfo{year}{2021}) \bibinfo{pages}{15242}.
\bibitem[{Boniou et~al.(2025)Boniou, Jay, Vinay, and Pierson}]{boniou2025revisiting}
\bibinfo{author}{V.~Boniou}, \bibinfo{author}{S.~Jay}, \bibinfo{author}{G.~Vinay}, \bibinfo{author}{J.-L. Pierson},
\newblock \bibinfo{title}{Revisiting the linear forcing of turbulence in two-phase flows},
\newblock \bibinfo{journal}{Physical Review Fluids} \bibinfo{volume}{10} (\bibinfo{year}{2025}) \bibinfo{pages}{024301}.
\bibitem[{Bassenne et~al.(2016)Bassenne, Urzay, Park, and Moin}]{bassenne2016constant}
\bibinfo{author}{M.~Bassenne}, \bibinfo{author}{J.~Urzay}, \bibinfo{author}{G.~I. Park}, \bibinfo{author}{P.~Moin},
\newblock \bibinfo{title}{Constant-energetics physical-space forcing methods for improved convergence to homogeneous-isotropic turbulence with application to particle-laden flows},
\newblock \bibinfo{journal}{Physics of Fluids} \bibinfo{volume}{28} (\bibinfo{year}{2016}).
\bibitem[{Jain(2022)}]{jain2022accurate}
\bibinfo{author}{S.~S. Jain},
\newblock \bibinfo{title}{Accurate conservative phase-field method for simulation of two-phase flows},
\newblock \bibinfo{journal}{J. Comput. Phys.} \bibinfo{volume}{469} (\bibinfo{year}{2022}) \bibinfo{pages}{111529}.
\bibitem[{Jain and Moin(2022)}]{jain2022kinetic}
\bibinfo{author}{S.~S. Jain}, \bibinfo{author}{P.~Moin},
\newblock \bibinfo{title}{A kinetic energy--and entropy-preserving scheme for compressible two-phase flows},
\newblock \bibinfo{journal}{Journal of Computational Physics} \bibinfo{volume}{464} (\bibinfo{year}{2022}) \bibinfo{pages}{111307}.
\bibitem[{Hwang and Jain(2024)}]{hwang2024robust}
\bibinfo{author}{H.~Hwang}, \bibinfo{author}{S.~S. Jain},
\newblock \bibinfo{title}{A robust phase-field method for two-phase flows on unstructured grids},
\newblock \bibinfo{journal}{Journal of Computational Physics} \bibinfo{volume}{507} (\bibinfo{year}{2024}) \bibinfo{pages}{112972}.
\bibitem[{Collis et~al.(2022)Collis, Mirjalili, Jain, and Mani}]{Collis2022}
\bibinfo{author}{H.~Collis}, \bibinfo{author}{S.~Mirjalili}, \bibinfo{author}{S.~Jain, S}, \bibinfo{author}{A.~Mani},
\newblock \bibinfo{title}{Assessment of weno and teno schemes for the four equation- compressible two-phase flow model with regularization term},
\newblock \bibinfo{journal}{Annual Research Briefs}  (\bibinfo{year}{2022}) \bibinfo{pages}{151--165}.
\bibitem[{Scapin et~al.(2022)Scapin, Shahmardi, Chan, Jain, Mirjalili, Pelanti, and Brandt}]{scapin2022mass}
\bibinfo{author}{N.~Scapin}, \bibinfo{author}{A.~Shahmardi}, \bibinfo{author}{W.~Chan}, \bibinfo{author}{S.~Jain}, \bibinfo{author}{S.~Mirjalili}, \bibinfo{author}{M.~Pelanti}, \bibinfo{author}{L.~Brandt},
\newblock \bibinfo{title}{A mass-conserving pressure-based method for two-phase flows with phase change},
\newblock in: \bibinfo{booktitle}{Center for Turbulence Research Proceedings of the Summer Program}.
\bibitem[{Brown et~al.(2023)Brown, Jain, and Moin}]{brown2023phase}
\bibinfo{author}{L.~Brown}, \bibinfo{author}{S.~Jain}, \bibinfo{author}{P.~Moin}, \bibinfo{title}{A Phase Field Model for Simulating the Freezing of Supercooled Liquid Droplets}, \bibinfo{type}{Technical Report}, SAE Technical Paper, \bibinfo{year}{2023}.
\bibitem[{Jain(2023)}]{jain_solsurf2023}
\bibinfo{author}{S.~Jain, S},
\newblock \bibinfo{title}{Modeling soluble surfactants in two-phase flows},
\newblock \bibinfo{journal}{Annual Research Briefs}  (\bibinfo{year}{2023}) \bibinfo{pages}{135--145}.
\bibitem[{Jain(2024)}]{jain2024model}
\bibinfo{author}{S.~S. Jain},
\newblock \bibinfo{title}{A model for transport of interface-confined scalars and insoluble surfactants in two-phase flows},
\newblock \bibinfo{journal}{Journal of Computational Physics} \bibinfo{volume}{515} (\bibinfo{year}{2024}) \bibinfo{pages}{113277}.
\bibitem[{Brackbill et~al.(1992)Brackbill, Kothe, and Zemach}]{Brackbill1992}
\bibinfo{author}{J.~Brackbill}, \bibinfo{author}{D.~Kothe}, \bibinfo{author}{C.~Zemach},
\newblock \bibinfo{title}{A continuum method for modeling surface tension},
\newblock \bibinfo{journal}{J. Comput. Phys.} \bibinfo{volume}{100} (\bibinfo{year}{1992}) \bibinfo{pages}{335--354}.
\bibitem[{Jain(2023)}]{jain2023implementation}
\bibinfo{author}{S.~S. Jain}, \bibinfo{title}{Implementation of the phase field method with the Immersed Boundary Method for application to wave energy converters}, Ph.D. thesis, Virginia Tech, \bibinfo{year}{2023}.
\bibitem[{Tryggvason et~al.(2011)Tryggvason, Scardovelli, and Zaleski}]{Tryggvason2011}
\bibinfo{author}{G.~Tryggvason}, \bibinfo{author}{R.~Scardovelli}, \bibinfo{author}{S.~Zaleski}, \bibinfo{title}{Direct numerical simulations of gas--liquid multiphase flows}, \bibinfo{publisher}{Cambridge University Press}, \bibinfo{year}{2011}.
\bibitem[{Sadek and Aluie(2018)}]{sadek2018extracting}
\bibinfo{author}{M.~Sadek}, \bibinfo{author}{H.~Aluie},
\newblock \bibinfo{title}{Extracting the spectrum of a flow by spatial filtering},
\newblock \bibinfo{journal}{Physical Review Fluids} \bibinfo{volume}{3} (\bibinfo{year}{2018}) \bibinfo{pages}{124610}.
\bibitem[{Aluie(2013)}]{aluie2013scale}
\bibinfo{author}{H.~Aluie},
\newblock \bibinfo{title}{Scale decomposition in compressible turbulence},
\newblock \bibinfo{journal}{Physica D: Nonlinear Phenomena} \bibinfo{volume}{247} (\bibinfo{year}{2013}) \bibinfo{pages}{54--65}.
\bibitem[{Zhao and Aluie(2018)}]{zhao2018inviscid}
\bibinfo{author}{D.~Zhao}, \bibinfo{author}{H.~Aluie},
\newblock \bibinfo{title}{Inviscid criterion for decomposing scales},
\newblock \bibinfo{journal}{Physical Review Fluids} \bibinfo{volume}{3} (\bibinfo{year}{2018}) \bibinfo{pages}{054603}.
\bibitem[{Freund and Ferrante(2019)}]{freund2019wavelet}
\bibinfo{author}{A.~Freund}, \bibinfo{author}{A.~Ferrante},
\newblock \bibinfo{title}{Wavelet-spectral analysis of droplet-laden isotropic turbulence},
\newblock \bibinfo{journal}{Journal of Fluid Mechanics} \bibinfo{volume}{875} (\bibinfo{year}{2019}) \bibinfo{pages}{914--928}.
\bibitem[{Jain and Elnahhas(2022)}]{jain2022modeling}
\bibinfo{author}{S.~Jain}, \bibinfo{author}{A.~Elnahhas},
\newblock \bibinfo{title}{Modeling of subgrid-scale interfacial area for turbulent two-phase flows},
\newblock \bibinfo{journal}{Bulletin of the American Physical Society} \bibinfo{volume}{67} (\bibinfo{year}{2022}).
\bibitem[{Deane and Dale~Stokes(2002)}]{Deane2002}
\bibinfo{author}{G.~Deane}, \bibinfo{author}{M.~Dale~Stokes},
\newblock \bibinfo{title}{Scale dependence of bubble creation mechanisms in breaking waves} \bibinfo{volume}{418} (\bibinfo{year}{2002}) \bibinfo{pages}{839--44}.
\bibitem[{Mukherjee et~al.(2019)Mukherjee, Safdari, Shardt, Kenjere{\v{s}}, and Van~den Akker}]{mukherjee2019droplet}
\bibinfo{author}{S.~Mukherjee}, \bibinfo{author}{A.~Safdari}, \bibinfo{author}{O.~Shardt}, \bibinfo{author}{S.~Kenjere{\v{s}}}, \bibinfo{author}{H.~E. Van~den Akker},
\newblock \bibinfo{title}{Droplet--turbulence interactions and quasi-equilibrium dynamics in turbulent emulsions},
\newblock \bibinfo{journal}{Journal of Fluid Mechanics} \bibinfo{volume}{878} (\bibinfo{year}{2019}) \bibinfo{pages}{221--276}.
\bibitem[{Roccon et~al.(2023)Roccon, Zonta, and Soldati}]{roccon2023phase}
\bibinfo{author}{A.~Roccon}, \bibinfo{author}{F.~Zonta}, \bibinfo{author}{A.~Soldati},
\newblock \bibinfo{title}{Phase-field modeling of complex interface dynamics in drop-laden turbulence},
\newblock \bibinfo{journal}{Physical Review Fluids} \bibinfo{volume}{8} (\bibinfo{year}{2023}) \bibinfo{pages}{090501}.
\bibitem[{Li et~al.(2017)Li, Miller, Wang, Koley, and Katz}]{li2017size}
\bibinfo{author}{C.~Li}, \bibinfo{author}{J.~Miller}, \bibinfo{author}{J.~Wang}, \bibinfo{author}{S.~Koley}, \bibinfo{author}{J.~Katz},
\newblock \bibinfo{title}{Size distribution and dispersion of droplets generated by impingement of breaking waves on oil slicks},
\newblock \bibinfo{journal}{Journal of Geophysical Research: Oceans} \bibinfo{volume}{122} (\bibinfo{year}{2017}) \bibinfo{pages}{7938--7957}.
\bibitem[{Jain and Elnahhas(2023)}]{jain_forced2023}
\bibinfo{author}{S.~Jain, S}, \bibinfo{author}{A.~Elnahhas},
\newblock \bibinfo{title}{Stationary states of forced two-phase turbulence},
\newblock \bibinfo{journal}{Annual Research Briefs}  (\bibinfo{year}{2023}) \bibinfo{pages}{119--133}.

\end{thebibliography}

\end{document}